\documentclass[12pt]{article}
\pdfoutput=1

\usepackage{draft} 
\usepackage{hyperref}
\usepackage{graphicx,color,subfig}
\usepackage{cite}
\usepackage{mciteplus}
\usepackage{skak}
\usepackage{empheq}
\DeclareFontFamily{OT1}{pzc}{}
\DeclareFontShape{OT1}{pzc}{m}{it}{<-> s * [1.10] pzcmi7t}{}
\DeclareMathAlphabet{\mathpzc}{OT1}{pzc}{m}{it}

\def\be#1\ee{\begin{align}#1\end{align}}

\begin{document}

\unitlength = .8mm

\begin{titlepage}

\begin{center}

\hfill \\
\hfill \\
\vskip 1cm

\title{The $c=1$ String Theory S-Matrix Revisited}

\author{Bruno Balthazar, Victor A. Rodriguez, Xi Yin}

\address{
Jefferson Physical Laboratory, Harvard University, \\
Cambridge, MA 02138 USA
}

\email{bbalthazar@g.harvard.edu, victorrodriguez@g.harvard.edu, xiyin@fas.harvard.edu}

\end{center}

\abstract{We revisit the perturbative S-matrix of $c=1$ string theory from the worldsheet perspective. We clarify the origin of the leg pole factors, the non-analyticity of the string amplitudes, and the validity as well as limitations of earlier computations based on resonance momenta. We compute the tree level 4-point amplitude and the genus one 2-point reflection amplitude by numerically integrating Virasoro conformal blocks with DOZZ structure constants on the sphere and on the torus, with sufficiently generic complex Liouville momenta, and find agreement with known answers from the $c=1$ matrix model. }

\vfill

\end{titlepage}

\eject

\begingroup
\hypersetup{linkcolor=black}
\tableofcontents
\endgroup

\section{Introduction} 

The two-dimensional noncritical ``$c=1$" string theory has been an invaluable source of inspiration in the exploration of string dualities and quantum gravity for nearly three decades (for reviews, see \cite{Klebanov:1991qa, Ginsparg:1993is, Jevicki:1993qn, Polchinski:1994mb, Martinec:2004td}). The perturbative excitations of the $c=1$ string theory are massless bosons in $1+1$ dimensions that interact and scatter off a Liouville wall. The full perturbative S-matrix (as well as its non-perturbative completion) was solved from the conjectured dual matrix quantum mechanics (the ``$c=1$ matrix model") in \cite{Moore:1991zv}. The string theory side of the story has been much murkier. The tree level string S-matrix was computed by an analytic continuation from ``resonance momenta" in \cite{DiFrancesco:1991ocm, DiFrancesco:1991daf}, where the relevant Liouville correlators were computed based on Coulomb gas integrals, and the result for the $1\to n$ amplitude has been successfully matched with the matrix model answer. This agreement was somewhat mysterious: firstly, the matrix model answer is only piecewise analytic in the momenta; secondly, the analogous resonance momenta computation for other amplitudes, such as the tree level $2\to 2$ scattering, fails to reproduce the anticipated answer from the matrix model. The exact solution of Liouville structure constants was subsequently discovered in \cite{Dorn:1994xn, Zamolodchikov:1995aa}, which in principle allows for explicit evaluation of Liouville correlators by integration Virasoro conformal blocks. This has been applied to the study of $c<1$ minimal string theory amplitudes, or ``minimal Liouville gravity", in \cite{Belavin:2003pu, Belavin:2005yj, Belavin:2006ex, Belavin:2008kv, Belavin:2010sr, Aleshkin:2016snp}. To the best of our knowledge, however, it has not been systematically applied to the study of the S-matrix in $c=1$ string theory.

In this paper, we will tie up a few loose ends on the tree level S-matrix in $c=1$ string theory, and extend the matching of the string theory and matrix model scattering amplitudes to genus one. Firstly, we explain the origin of the so-called ``leg pole factors" as simply a result of proper normalization of physical vertex operators in the $c=25$ Liouville theory on the worldsheet of the string. After formulating the $c=1$ string amplitude as an integral of Virasoro conformal blocks, we will understand in detail the analytic property of the amplitude in the Liouville momenta, and recover the known piecewise-analytic structure of the S-matrix of the $c=1$ matrix model. Indeed, the analytic continuation to resonance momenta reduces the Liouville correlator on the sphere to that of the linear dilaton CFT, multiplied by a normalization constant that may or may not diverge. Based on this, we will explain precisely why the resonance computation produces the correct tree level $1\to 3$ amplitude but not the $2\to 2$ amplitude.

To explicitly compute the $c=1$ string amplitude for general momenta, we resort to the numerical approach. First, we evaluate the Liouville correlator by integrating the Virasoro conformal blocks multiplied by DOZZ structure constants over the internal Liouville momenta. We then numerically integrate the Liouville correlators along with the $c=1$ ``matter" and ghost contributions to produce the string amplitude. For the tree level 4-point amplitude, a key ingredient that allows for efficient computation is Zamolodchikov's recursive representation of the sphere 4-point Virasoro conformal block \cite{Zamolodchikov:1985ie, 1987TMP....73.1088Z}. A generalization of such recursive representations to general Virasoro conformal blocks was found recently in \cite{Cho:2017oxl}. We will make use of two special cases of the formulae presented in \cite{Cho:2017oxl} for torus 2-point blocks, in the OPE channel and in the necklace channel. These will be used in the numerical evaluation of the Liouville torus 2-point function of the genus one contribution to the $1\to 1$ reflection amplitude in $c=1$ string theory.

The $1\to n$ S-matrix of $c=1$ string theory takes the form \cite{Ginsparg:1993is}
\ie{}
&S_{1\to n}(\omega; \omega_1,\cdots, \omega_n) = \delta\left(\omega - \sum_{i=1}^n \omega_i\right) {\cal A}_{1\to n}(\omega_1,\cdots,\omega_n),
\fe
where $\omega$ is the energy of the incoming massless particle, $\omega_1,\cdots,\omega_n$ are the energies of the outgoing particles (after reflection from the Liouville wall). It is subject to the unitarity condition
\ie
\int_{\omega_i\geq 0,~\sum_{j=1}^n\omega_j=\omega} \prod_{i=1}^n d\omega_i \, {\left| {\cal A}_{1\to n}(\omega_1,\cdots, \omega_n) \right|^2 \over \omega\, \omega_1\cdots\omega_n} = 1
\fe
for every $\omega>0$.
${\cal A}_{1\to n}$ has a perturbative expansion
\ie{}
&{\cal A}_{1\to n}(\omega_1,\cdots,\omega_n)=
\sum_{L=0}^\infty g^{n-1+2L} {\cal A}_{1\to n}^{(L)}(\omega_1,\cdots,\omega_n),
\fe
where $g$ is the string coupling, and ${\cal A}^{(L)}$ represents the genus $L$ string amplitude. 
From the matrix model, one anticipates the answer
\ie\label{matrixanswer}{}
& {\cal A}_{1\to 1}^{(0)} = \omega,~~~ {\cal A}_{1\to 2}^{(0)} =  i \omega_1\omega_2 \omega,~~~ {\cal A}_{1\to 3}^{(0)} = i \omega_1 \omega_2\omega_3 \omega \left( 1 + i \omega \right),~~~\cdots
\\
& {\cal A}_{1\to 1}^{(1)} = {1\over 24} \left( i \omega^2 + 2i \omega^4 - \omega^5 \right), ~~~ \cdots
\fe
where we have written $\omega = \sum_{i=1}^n \omega_i$. We will reproduce the above results for ${\cal A}_{1\to 3}^{(0)}$ and ${\cal A}_{1\to 1}^{(1)}$ from the string worldsheet by numerically integrating conformal blocks. Note that the real part of ${\cal A}_{1\to 1}^{(1)}$ is fixed by ${\cal A}_{1\to 2}^{(0)}$ through perturbative unitarity. The agreement on the imaginary part of ${\cal A}_{1\to 1}^{(1)}$ is the first nontrivial test of the equivalence of the string S-matrix with that of the matrix duality beyond the tree level S-matrix.

Let us note that the S-matrix of massless particles in 1+1 spacetime dimensions is extremely subtle. Usually in quantum field theory, in and out asymptotic states with real physical momenta are defined as limits of far separated wave packets; such a separation of wave packets is not available here. In \cite{Moore:1991zv, Ginsparg:1993is}, the S-matrix as computed from the matrix model is essentially defined through analytic continuation from complex momenta. This is also technically necessary from the string worldsheet, in order to ensure the convergence of the integration over moduli space of punctured Riemann surfaces. However, if we take the real momenta limit, say $\omega_i\in \mathbb{R}+i\epsilon_i$ with $\epsilon_i\to 0$, the result is generally sensitive to the ordering of $\epsilon_i$. For instance, the tree level $2\to 2$ amplitude takes the form\footnote{This is consistent with the result obtained from Euclidean Green's function in \cite{Moore:1991zv, Ginsparg:1993is} for purely imaginary momenta, but differs from some assertions in the literature for the Lorentzian amplitude (e.g. in \cite{Polchinski:1991uq, Klebanov:1991qa, Ginsparg:1993is,Mandal:1991ua}) where ${\rm max}\{\omega_j\}$ for real $\omega_j$ was written in place of ${\rm Imax}\{\omega_j\}$ in (\ref{matrix22}). We believe this is due to ambiguities in the definition of asymptotic states of multiple massless particles in $c=1$ string theory. }
\ie\label{matrix22}
{\cal A}_{2\to 2}^{(0)} = i\omega_1\omega_2\omega_3\omega_4(1+i \,{\rm Imax} \{\omega_j\} ),
\fe
where $\omega_1,\cdots , \omega_4$ are the energies of the incoming and outgoing particles, all of which are taken to have positive real {\it and} positive imaginary parts, and ${\rm Imax} \{\omega_j\}$ is defined to be the element with the largest {\it imaginary part} among $\{\omega_1, \omega_2, \omega_3, \omega_4\}$. The non-analytic feature of (\ref{matrix22}) is in fact due to intermediate on-shell particles and is required by unitarity. We will show that ${\cal A}_{2\to 2}^{(0)}$ is nonetheless related to ${\cal A}_{1\to 3}^{(0)}$ by a slightly unconventional crossing relation. We will explain the validity of this analytic continuation from the structure of the conformal block integral in the worldsheet computation. 

The paper is organized as follows. In section 2 we review the general structure of the S-matrix in $c=1$ string theory, and summarize the results from the matrix model side. The computation of the $c=1$ string amplitudes on the sphere and torus will be discussed in section 3 and 4 respectively. We conclude in section 5 with some future perspectives. 

\section{The structure of the perturbative S-matrix in $c=1$ string theory}

\subsection{The worldsheet theory}

The worldsheet formulation of the $c=1$ string perturbation theory is based on a time-like free boson $X^0$ together with $c=25$ Liouville theory and the $b,c$ ghost system. The Liouville CFT is governed by the action 
\ie
S_L = {1\over 4\pi} \int d^2z \sqrt{g} \left( g^{mn} \partial_m \phi \partial_n \phi + Q R \phi + 4\pi \mu e^{2b\phi} \right),
\fe 
where the central charge is related to the background charge $Q=b+b^{-1}$ by $c=1+6Q^2$. The case of interest $c=25$ corresponds to $b=1$. The Virasoro primaries of Liouville CFT are scalar operators $V_P$ labeled by the ``Liouville momentum" $P\in\mathbb{R}_{\geq 0}$. $V_P$ has scaling dimension $\Delta = h+\tilde h = {Q^2\over 2} + 2P^2$. Our normalization convention is such that, in the $\phi\to -\infty$ limit (where the Liouville potential vanishes), $V_P$ takes the form\footnote{In the literature, such a vertex operator in Liouville theory would often be written as an ``exponential operator" $S(P)^{-{1\over 2}} e^{(Q+2iP)\phi}$, by analogy with linear dilaton CFT. We find this notation unnecessarily misleading, and will simply use the notation $V_P$ instead. }
\ie\label{normc}
V_P \sim S(P)^{-{1\over 2}} e^{(Q+2iP)\phi} + S(P)^{{1\over 2}} e^{(Q-2iP)\phi}.
\fe 
Here $S(P)$ is the reflection phase
\ie\label{phase}
S(P) = (\pi \mu \gamma(b^2))^{-2iP\over b} {\gamma( {2iP\over b}) \over b^2 \gamma(-2iPb)},
\fe
where $\gamma(x)\equiv \Gamma(x)/\Gamma(1-x)$.
With (\ref{normc}), the vertex operators $V_P$ are delta-function normalized, i.e. their 2-point functions are
\ie\label{Liouville2pt}
\left\langle V_{P_1}(z,\bar z) V_{P_2}(0) \right\rangle = \pi{\delta(P_1-P_2)\over |z|^{\Delta_1+\Delta_2}}.
\fe
The 3-point functions are given by the DOZZ structure constants \cite{Dorn:1994xn, Zamolodchikov:1995aa}. Following the convention of Appendix B of \cite{Collier:2017shs}, we have
\ie\label{dozz}
& \left\langle V_{P_1}(z_1,\bar z_1)  V_{P_2}(z_2,\bar z_2)V_{P_3}(z_3,\bar z_3) \right\rangle = \left[ \pi \mu \gamma(b^2) b^{2-2b^2} \right]^{-{Q\over 2b}}  {{\cal C}(P_1, P_2, P_3)\over |z_{12}|^{\Delta_1 + \Delta_2 - \Delta_3} |z_{23}|^{\Delta_2 + \Delta_3 - \Delta_1} |z_{31}|^{\Delta_3 + \Delta_1 - \Delta_2}},
\fe
where
\ie\label{cpppq}
& {\cal C} (P_1, P_2, P_3) = {\Upsilon_b'(0) \over \Upsilon_b({Q\over 2} +i(P_1+P_2+P_3)) }
\left[ {(\Upsilon_b(2iP_1) \Upsilon_b(-2iP_1) )^{1\over 2}\over  \Upsilon_b({Q\over 2} + i(P_2+P_3-P_1)) } \times (2~{\rm permutations})\right].
\fe
$\Upsilon_b(x)$ is the Barnes double Gamma function, defined by (the analytic continuation of)
\ie
\log\Upsilon_b(x) = \int_0^\infty {dt\over t} \left[ \left( {Q\over 2}-x \right)^2 e^{-t} - {\sinh^2\left[\left({Q\over 2}-x\right){t\over 2} \right]\over \sinh{tb\over 2}\sinh{t\over 2b} } \right],~~~ 0<{\rm Re} (x) < Q.
\fe
It has the useful properties
\ie
& \Upsilon_b(Q-x) = \Upsilon_b(x),
\\
& \Upsilon_b(x+b) = \gamma(bx) b^{1-2bx} \Upsilon_b(x),
\\
& \Upsilon_b(x+b^{-1}) = \gamma(b^{-1} x) b^{{2x\over b}-1} \Upsilon_b(x).
\fe
Furthermore, $\Upsilon_b(x)$ is an entire analytic function with simple zeroes at $x=mb + n/b$, for integers $m,n\leq 0$ or $m,n\geq 1$.

The $c=25$ Liouville CFT of interest is obtained by taking the $b\to 1$ limit. 
In this limit, the prefactor $\left[ \pi \mu \gamma(b^2) b^{2-2b^2} \right]^{-{Q\over 2b}}$ in the 3-point function (\ref{dozz}) looks singular, but it can be absorbed by a rescaling of the Liouville cosmological constant $\mu$, which amounts to a renormalization of the string coupling. With this understanding, we will drop this prefactor, and write (\ref{cpppq}) as the 3-point function coefficient of the Liouville primaries. In the $b\to 1$ limit, it becomes
\ie\label{cppp}
{\cal C}(P_1, P_2, P_3) = {1\over \Upsilon_1(1+i(P_1+P_2+P_3)) } \left[ { 2P_1 \Upsilon_1(1+2iP_1)\over  \Upsilon_1(1 + i(P_2+P_3-P_1)) } \times (2~{\rm permutations}) \right].
\fe
An interesting property that will be useful later is that, under analytic continuation of $P_i\to -P_i$, ${\cal C}(P_1, P_2, P_3)$ flips sign.

The BRST cohomology classes corresponding to 1-particle asymptotic states are represented by vertex operators of the form
\ie\label{ima}
{\cal V}_\omega^\pm = g_s :\! e^{\pm i\omega X^0}\!\!: V_{P={\omega\over 2}}.
\fe
Here ${\cal V}_\omega^+$ represents an incoming mode and ${\cal V}_\omega^-$ an outgoing mode, of energy $\omega>0$. The Liouville momentum $P$ is identified with $\omega\over 2$, so that (\ref{ima}) is a weight $(1,1)$ Virasoro primary. The perturbative string amplitudes are formulated as in the usual bosonic string theory. The simplest nontrivial example is the tree level $1\to 2$ scattering amplitude, computed by
\ie
\langle c\tilde c {\cal V}_\omega^+(z_1, \bar z_1) c\tilde c{\cal V}_{\omega_1}^-(z_2,\bar z_2) c\tilde c{\cal V}_{\omega_2}^-(z_3, \bar z_3) \rangle = ig_s^3 C_{S^2} \delta(\omega-\omega_1-\omega_2) {\cal C}\left({\omega\over 2}, {\omega_1\over 2}, {\omega_2\over 2}\right).
\fe
The formula (\ref{cppp}) drastically simplifies in the special case $P_3=P_1+P_2$, giving the result (using $\Upsilon_1(1)=1$)
\ie
{\cal C}\left({\omega\over 2}, {\omega_1\over 2}, {\omega_2\over 2}\right) = \omega \omega_1 \omega_2.
\fe
Thus, we recover the matrix model answer for ${\cal A}_{1\to 2}^{(0)}$ in (\ref{matrixanswer}) (with the identification $g_s^3 C_{S^2} = g$). In the literature, this is a well known result, but it is often stated with an explicit inclusion of ``leg pole factors" \cite{Klebanov:1991qa, Ginsparg:1993is, Polchinski:1994mb}. Here we see that the leg pole factor is already taken into account due to the normalization factor $S(P)^{-{1\over 2}}$ in the definition of $V_P$, which was needed to normalize the 2-point function of Liouville primaries.\footnote{The nature of the leg pole factor as a scattering phase was already pointed out in \cite{Klebanov:1991qa}.}

\subsection{The dual matrix model}

Before discussing the more general $c=1$ string amplitudes from the worldsheet perspective, let us briefly review the anticipated answer from the dual $c=1$ matrix model. The $c=1$ matrix model is defined as a suitable $N\to \infty$ limit of the $U(N)$ gauged matrix quantum mechanics with the Hamiltonian $H = {1\over 2} {\rm Tr} (P^2-X^2)$, where $X$ is a Hermitian $N\times N$ matrix on which the $U(N)$ gauge symmetry acts by the adjoint representation, and $P$ is the canonically conjugate momentum matrix. A commonly used equivalent formulation is the system of free fermions subject to the Hamiltonian $H={1\over 2}(p^2-x^2)$, filling the region $x>\sqrt{p^2+2\mu}$ in the phase space, for some $\mu>0$. This configuration is perturbatively stable, which suffices for our discussion.\footnote{There are various non-perturbative completions of the $c=1$ matrix model. One particularly interesting and extensively studied version is the type 0B matrix model \cite{Takayanagi:2003sm, Douglas:2003up}.}

The quasi-particles of the $c=1$ matrix model that are dual to the asymptotic states of the $c=1$ string theory are collective excitations of the fermi surface. In the asymptotic region $x\to \infty$, collective excitations of the fermi surface travels exponentially fast in the $x$ coordinate. To identify them with the massless scalars in the $\phi\to -\infty$ region of the $c=1$ string theory, one expects roughly an exponential map between the $x$ coordinate of the fermi sea excitation and the Liouville coordinate of the string mode. Detailed descriptions of the collective excitations are given in \cite{Sengupta:1990bt, Das:1990kaa, Polchinski:1991uq, Moore:1991zv, Moore:1992gb, Polchinski:1994mb, Gross:1990st}. Here we follow the approach of \cite{Moore:1991zv, Moore:1992gb}, where the S-matrix of collective fields are extracted from the LSZ limit of the Green functions of fermion density operators. Essentially, each asymptotic particle is traded with a pair of fermion creation and annihilated operators, which are then contracted using the free fermion Green function which includes a reflection factor
\ie
R(\omega) = i \mu^{i\omega} \left[ {(1+i e^{-\pi\omega})\Gamma({1\over 2} - i\omega) \over (1-i e^{-\pi\omega})\Gamma({1\over 2} + i\omega)} \right]^{1\over 2}.
\fe
For instance, the exact $1\to 1$ and $1\to 2$ S-matrix elements are given by
\ie\label{aam}
& {\cal A}_{1\to 1}(\omega) = \int_0^\omega dx\, R^*(\mu- x) R(\mu + \omega - x),
\\
& {\cal A}_{1\to 2} (\omega_1, \omega_2) = \int_0^{\omega_2} dx R^*(\mu - x) R(\mu + \omega- x) - \int_{\omega_1}^\omega dx \, R^*(\mu - x) R(\mu + \omega - x),~~~\omega=\omega_1+\omega_2.
\fe
Expanding (\ref{aam}) perturbatively in $1/\mu$, with the identification $g=\mu^{-1}$, one produces (\ref{matrixanswer}).

\section{Tree level 4-point string amplitude from Liouville correlator}

\subsection{The $1\to 3$ amplitude}

We begin with the tree level $1\to 3$ amplitude in $c=1$ string theory, 
\ie\label{amp4pt}
&\int d^2 z \left\langle {\cal V}_\omega^+(z,\bar z) {\cal V}_{\omega_1}^-(0){\cal V}_{\omega_2}^-(1){\cal V}_{\omega_3}^-(\infty) \right\rangle
\\
&= i g_s^4 C_{S^2} \delta\left(\omega- \sum_{j=1}^3\omega_j\right) \int d^2z |z|^{\omega\omega_1}|1-z|^{\omega\omega_2} \left\langle V_{\omega\over 2}(z,\bar z) V_{\omega_1\over 2}(0) V_{\omega_2\over 2}(1) V_{\omega_3\over 2}(\infty) \right\rangle_{\rm Liouville}.
\fe
The Liouville 4-point function is given by 
\ie\label{liouville4pt}
& \left\langle V_{\omega\over 2}(z,\bar z) V_{\omega_1\over 2}(0) V_{\omega_2\over 2}(1) V_{\omega_3\over 2}(\infty) \right\rangle_{\rm Liouville} 
\\
& = \int_0^\infty {dP\over\pi} \, {\cal C}({\omega\over 2}, {\omega_1\over 2}, P) {\cal C}({\omega_2\over 2}, {\omega_3\over 2}, P) 
F_P(z) F_P(\bar z),
\fe
where ${\cal C}$ is the Liouville structure constant, as given by (\ref{cppp}) with our normalization of vertex operators. $F_P(z)$ is the holomorphic conformal block with internal weight corresponding to Liouville momentum $P$, namely
\ie
F_P(z) = {\cal F}(1+{\omega^2\over 4}, 1+{\omega_1^2\over 4}, 1+{\omega_2^2\over 4}, 1+{\omega_3^2\over 4}; 1+P^2 |z)
\fe
where ${\cal F}(h_1, h_2, h_3, h_3; h|z)$ is the sphere 4-point $c=25$ Virasoro conformal block, with external weights $h_1, \cdots, h_4$ and internal weight $h$.

Naively, there is an immediate puzzle in comparison to the anticipated matrix model answer: the matrix model amplitude ${\cal A}_{1\to 3}^{(0)}$ as in (\ref{matrixanswer}) for real energies has both real and imaginary parts, while the $z$-integral in (\ref{amp4pt}) is formally real for real $\omega$'s, and it would seem that the two cannot possibly agree. The $z$-integral is a priori divergent and must be regularized. We will see in section \ref{regz} that the regularized $z$-integral nonetheless fails to converge for strictly real $\omega$'s, leading to a non-analyticity of the string amplitude (\ref{amp4pt}) at real $\omega$'s. The correct computation of the string amplitude requires a suitable $i\epsilon$ prescription for the vertex operators representing in and out states, which amounts to giving imaginary parts to the $\omega$'s in (\ref{amp4pt}). We will see that such a prescription will indeed produce a string amplitude that agrees with the matrix model.

The integrand of (\ref{liouville4pt}) is analytic in $\omega, \omega_1, \omega_2, \omega_3$ apart from poles coming from the structure constants. Thus, the amplitude ${\cal A}_{1\to 3}^{(0)}$ given by the integral on the RHS of (\ref{amp4pt}) can be analytically continued, modulo the possibility of poles in $P$ crossing the $P$-integration contour in (\ref{liouville4pt}), and the regularization prescription for the $z$-integration near $z=0$, 1, and $\infty$. We further note that {\it integrand} of (\ref{liouville4pt}) changes sign if we analytically continue either $\omega$ or one of the $\omega_i$'s to minus itself. This gives rise to the possibility of a certain crossing symmetry for scattering amplitudes, to be discussed in section \ref{sec: 2to2}.

\subsubsection{Analytic continuation of Liouville correlator}
\label{sec:analytic_tree}

Let us first consider the analytic continuation of the Liouville correlator (\ref{liouville4pt}). The integrand in $P$ has poles located at
\ie\label{poles}
& P = \A + i n,~~~ n=\pm 1, \pm 2, \pm 3, \cdots
\\
&~~~~{\rm where}~ \A = {\pm\omega\pm\omega_1\over 2} ~{\rm and}~{\pm \omega_2 \pm \omega_3\over 2}
\fe
from the structure constants.\footnote{The conformal block has poles at $P=in/2$ for integer $n\geq 2$, with multiplicity $2(n-1)$ (from both the holomorphic and the anti-holomorphic parts), but they are canceled by the zeroes of the structure constants of multiplicity $2n$.} The integral is analytic in $\omega$ and $\omega_i$ provided that the poles (\ref{poles}) do not cross the $P$-contour. When some of the poles approach the $P$-contour, we can deform the contour appropriately to avoid the poles, until the contour is pinched by a pair of poles. For instance, maintaining $\omega = \omega_1+ \omega_2+\omega_3$, we may analytically continue in $\omega$ by giving it an imaginary part, and deform the $P$-contour if necessary to avoid the poles. As we take $\omega(=\omega_1+\omega_2+\omega_3)\to 2i$, the pair of poles of ${\cal C}({\omega\over 2}, {\omega_1\over 2},P)$ at
\ie
P = {\omega+\omega_1\over 2} - i~~~{\rm and}~~ {-\omega+\omega_1\over 2}+i
\fe
pinch the $P$-contour at $P={\omega_1\over 2}$. There is a further subtlety in this limit: the numerator of ${\cal C}({\omega\over 2}, {\omega_1\over 2},P)$ as given by (\ref{cppp}) has a double zero at $\omega=2i$; on the other hand, there is another pole of ${\cal C}({\omega_2\over 2}, {\omega_3\over 2},P)$ at $P=-{\omega_2+\omega_3\over 2}+i$ that also approaches ${\omega_1\over 2}$ in the $\omega\to 2i$ limit. Their net effect is such that, in the $\omega\to 2i$ limit, the contributions from the $P$-integral away from $P={\omega_1\over 2}$ vanishes, while a finite residue contribution at $P={\omega_1\over 2}$ survives. This residue contribution is proportional to a single conformal block with external weights 0, $1+{\omega_i^2\over 4}$ $(i=1,2,3)$, and internal weight $1+{\omega_1^2\over 4}$. This conformal block is simply equal to 1, and the Liouville correlator reduces to a 3-point function, namely
\ie
& \lim_{\omega = \omega_1+\omega_2+\omega_3\to 2i}\left\langle V_{\omega\over 2}(z,\bar z) V_{\omega_1\over 2}(0) V_{\omega_2\over 2}(1) V_{\omega_3\over 2}(\infty) \right\rangle_{\rm Liouville} 
\\
& = \lim_{\omega = \omega_1+\omega_2+\omega_3\to 2i} (-2i ) \,{\rm Res}_{P\to {\omega+\omega_1\over 2}-i} {\cal C}({\omega\over 2}, {\omega_1\over 2}, P) {\cal C}({\omega_2\over 2}, {\omega_3\over 2}, P) 
F_P(z) F_P(\bar z)
\\
&= -4 \prod_{j=1}^3 {\Gamma(1-i\omega_j)\over \Gamma(1+i\omega_j)}.
\fe
This is a special case of resonance momenta, where the Liouville correlator reduces to that of the corresponding linear dilaton CFT. While this analytic continuation is correct at the level of the Liouville correlator, it is not immediately obvious that it is compatible with the regularization needed in defining the moduli integral (\ref{amp4pt}). If we naively analytically continue the moduli space integrand in (\ref{amp4pt}) to $\omega=2i$, the resulting amplitude is
\ie
\left. g^2 {\cal A}_{1\to 3}^{(0)}\right|_{\omega_1+\omega_2+\omega_3 = 2i} = - 4i g_s^4 C_{S^2} \prod_{j=1}^3 {\Gamma(1-i\omega_j)\over \Gamma(1+i\omega_j)}\int d^2z\, |z|^{2i\omega_1} |1-z|^{2i\omega_2} = 4\pi g_s^4 C_{S^2} \omega_1\omega_2\omega_3
\fe
While not immediately obvious (due to the need of regularizing the $z$-integral), it will be justifed shortly that this is in fact the correct analytic continuation of the tree level $1\to 3$ amplitude from the physical domain. For now, let us note that the result agrees with that of the matrix model (\ref{matrixanswer}) for ${\cal A}_{1\to 3}^{(0)}$ when the latter is analytically continued to $\omega=2i$, provided that we identify $2\pi g_s^4 C_{S^2}=g^2$. Combining with the 3-point amplitude, we can fix the normalization
\ie
g = 2\pi g_s, ~~~ C_{S^2} = {2\pi \over g_s^2}.
\label{eq:norms1}
\fe

\subsubsection{Regularization of moduli integral}\label{regz}

Let us now turn to the regularization of the $z$-integral in (\ref{amp4pt}). The latter has potential divergences from the vicinity of $z=0$, 1, and $\infty$. In higher dimensional string theories, such divergences are regularized easily by analytic continuation in the momenta. The special kinematics of the $1+1$ dimensional S-matrix in $c=1$ string theory makes the analytic continuation in momenta rather subtle, even for tree level amplitudes. We will regularize the $z$-integral by explicitly subtracting off counter terms from the integrands, that is compatible with analyticity in the momenta. The basic idea is very simple: consider for instance the integral 
\ie
\int d^2z |z|^{x-2} \theta(1-|z|^2) = {2\pi\over x}
\fe
for ${\rm Re}(x)>0$. Its analytic continuation in $x$ can be obtained by adding a ``counter term", namely,
\ie
\int d^2z |z|^{x-2} \left[\theta(1-|z|^2) - 1 \right] = {2\pi\over x}
\fe
now holds for ${\rm Re}(x)<0$.

To regularize the string amplitude, consider the vicinity of $z=0$, where we can write the $z$-integral using the OPE of Liouville vertex operators in the form
\ie\label{zope}
\int d^2z \int_0^\infty {dP\over \pi} {\cal C}({\omega\over 2}, {\omega_1\over 2}, P) {\cal C}({\omega_2\over 2}, {\omega_3\over 2}, P)  |z|^{-2 -{1\over 2}(\omega-\omega_1)^2 + 2P^2} \sum_{n,m=0}^\infty a_{n,m} z^n \bar z^m ,
\fe
where $a_{0,0}=1$, and $a_{n,m}$'s are rational functions of internal and external weights, and thus are rational functions in $\omega$, $\omega_i$, and $P$. The $z$-integral is a priori divergent for $P^2< {1\over 4}{\rm Re}((\omega-\omega_1)^2)$. In particular, this divergence is always present in the physical domain, where $\omega$ and $\omega_1$ are real. We regularize this divergence by introducing a counter term in the $z$-integrand, that amounts to removing the terms proportional to $a_{n,m}$ in (\ref{zope}) with 
\ie
n=m < {1\over 4} {\rm Re}((\omega-\omega_1)^2) - P^2.
\fe
This ``$s$-channel" counter term for the integrand is
\ie
R_s = \sum_{0\leq n \leq {1\over 4} {\rm Re}((\omega-\omega_1)^2)} a_{n,n} \int_0^{\sqrt{{1\over 4} {\rm Re}((\omega-\omega_1)^2)-n}} {dP\over \pi} {\cal C}({\omega\over 2}, {\omega_1\over 2}, P) {\cal C}({\omega_2\over 2}, {\omega_3\over 2}, P)  |z|^{-2 -{1\over 2}(\omega-\omega_1)^2 + 2P^2+2n} .
\label{reg}
\fe
We will also need to include counter terms for the $t$ and $u$ channel OPEs
\ie
R_t = \left. R_s\right|_{z\to z-1, \,\omega_1\leftrightarrow\omega_2},~~~ R_u = |z|^{-4} \left(\left. R_s\right|_{z\to {1/ z}, \,\omega_1\leftrightarrow\omega_3}\right).
\fe
The full regularized $z$-integral for the $1\to 3$ amplitude is
\ie\label{reged}
\int d^2z \left[ |z|^{\omega\omega_1}|1-z|^{\omega\omega_2} \left\langle V_{\omega\over 2}(z,\bar z) V_{\omega_1\over 2}(0) V_{\omega_2\over 2}(1) V_{\omega_3\over 2}(\infty) \right\rangle_{\rm Liouville} - R_s - R_t - R_u\right].
\fe

Importantly, the string amplitude should be computed with an $i\epsilon$ prescription that assigns positive imaginary parts to the physical energies of incoming and outgoing asymptotic states. Our regularization preserves analyticity of the amplitude in the momenta, provided that ${\rm Im}((\omega-\omega_i)^2)$ do not cross zero where ${\rm Re}((\omega-\omega_i)^2)>0$, for $i=1,2,3$. This analyticity criterion is equivalent to ${\rm Im}(\omega-\omega_i)\not=0$. When ${\rm Im}(\omega-\omega_i)$ approaches zero, the regulated $z$-integrand contains a term of the form
\ie
\int_{P_*}^\infty dP\, f(P) |z|^{-2 + 2(P^2-P_*^2)} \sim -{f(P_*) \over 4P_* |z|^2\log|z|}
\fe
near $z=0$ (or an analogous expression related by crossing near $z=1$ or $\infty$), for some $P_*>0$, which leads to a $\log\log$ divergence in the $z$-integral. This is the source of the non-analytic behavior of the amplitude. Such non-analyticity does not show up in the tree level $1\to 3$ amplitude in the physics domain, but it will affect the $2\to 2$ amplitude and in fact leads to ambiguities in the physical domain, as we now discuss.

\subsection{The $2\to 2$ amplitude and crossing symmetry}
\label{sec: 2to2}

The tree level $2\to 2$ amplitude is given by
\ie\label{amp22}
&\int d^2 z \left\langle {\cal V}_{\omega_1}^+(z,\bar z) {\cal V}_{\omega_2}^+(0){\cal V}_{\omega_3}^-(1){\cal V}_{\omega_4}^-(\infty) \right\rangle
\fe
with the same regularization prescription as above. Formally, it can be obtained as {\it minus} the $1\to 3$ amplitude with one of the outgoing energies $\omega_i$ analytically continued to minus itself. Namely, 
\ie\label{anacon}
{\cal A}^{(0)}(\{\omega_1,\omega_2\}\to \{\omega_3,\omega_4\}) = - {\cal A}^{(0)}(\{\omega_1\}\to \{ -\omega_2, \omega_3,\omega_4\}),
\fe
provided that the continuation $\omega_2\to -\omega_2$ can be achieved while maintaining 
\ie\label{ieps}
{\rm Im}(\omega_1+\omega_2), ~{\rm Im}(\omega_1-\omega_3), ~ {\rm Im}(\omega_1-\omega_4) > 0.
\fe
The rather unusual minus sign on the RHS of (\ref{anacon}) is a consequence of the analytic property of Liouville structure constants (\ref{cppp}). This analytic continuation is indeed possible when ${\rm Im}(\omega_1)$ is the largest among $\{{\rm Im}(\omega_j), j=1,2,3,4\}$ (all of which are taken to be positive), and leads to the result (\ref{matrix22}).
 
Note that (\ref{matrix22}) is ambiguous in the real momenta limit, as the answer is sensitive to the ordering of the imaginary parts of $\omega_j$. This is due to ambiguities in the definition of asymptotic states of massless particles in 1+1 dimensions. 

In fact, we can understand the non-analyticity in the amplitude as follows. As a function of complex energies, (\ref{matrix22}) is non-analytic along the locus $\omega_i - \omega_j\in\mathbb{R}$ for some $\omega_i$ labeling the energy of an incoming particle and $\omega_j$ that of an outgoing particle. This occurs precisely when an intermediate particle goes on-shell. For instance, if $\omega_1 - \omega_3\in \mathbb{R}_{>0}$, a particle propagates with real positive energy in the $t$-channel. The discontinuity of ${\cal A}_{2\to 2}^{(0)}$ as ${\rm Im}(\omega_1-\omega_3)$ changes from negative to positive across zero is given by
\ie\label{discatt}
{\rm disc} {\cal A}_{2\to 2}^{(0)} = \omega_1 \omega_2 \omega_3 \omega_4 (\omega_1 - \omega_3).
\fe
It is generally expected from unitarity that such a discontinuity should be captured by the factorized amplitude through the intermediate on-shell particle. Indeed, (\ref{discatt}) precisely agrees with the discontinuity of
\ie
\int_0^\infty\frac{d\omega}{2\pi}\frac{i}{\omega}\frac{\mathcal{A}_{1\to 2}^{(0)}(\omega_1\to\{\omega,\omega_3\})\mathcal{A}_{2\to 1}^{(0)} (\{\omega_2,\omega\}\to\omega_4)}{\omega_1-\omega_3-\omega},
\fe
as $\omega_1 - \omega_3$ crosses the positive real axis.

\subsubsection{Resonance momenta}

Let us comment on the analytic continuation of the $2\to 2$ amplitude to resonance momenta, which amounts to taking $\omega_1 + \omega_2 = \omega_3 + \omega_4 = 2i(1-\epsilon)$, and send $\epsilon\to 0^+$. Writing the Liouville correlator via its conformal block decomposition in the $12\to 34$ channel, provided that we keep ${\rm Im}(\omega_i)>0$ for all $i$, no poles would have crossed the contour and thus the $P$-integration contour remains along the real axis. In the $\epsilon\to 0^+$ limit, the double poles of the structure constants at $P=\pm {\omega_1+\omega_2\over 2}\mp i=\pm {\omega_3+\omega_4\over 2}\mp i$ pinch the contour at $P=0$. There is also a double zero at $P=0$ from the structure constants. The net result is that the Liouville 4-point function in the $\epsilon\to 0^+$ limit is dominated by the contribution from the contour integral near $P=0$, giving the linear dilaton 4-point function up to a normalization factor that diverges like $1/\epsilon$.

This does {\it not} imply that the $2\to 2$ string tree amplitude at the resonance momenta can be computed from the linear dilaton correlator, however. Firstly, the moduli $z$-integral of the linear dilaton 4-point function together with the time-like free boson contribution vanishes like $\epsilon$ in the resonance limit (this vanishing was remarked in \cite{Ginsparg:1993is}, for instance). This cancels the above mentioned $1/\epsilon$ divergence and would produce a finite contribution. However, we cannot ignore the contribution from the rest of the $P$-integral, which, unlike the $1\to 3$ resonance amplitude considered in section \ref{sec:analytic_tree}, does not vanish in the $\epsilon\to 0^+$ limit. Thus, even after analytic continuation to resonance momenta, the $2\to 2$ amplitude does {\it not} reduce to the contribution from a single Virasoro conformal block.

\subsection{Numerical results}

We now compute numerically the tree level $1\to 3$ amplitude in $c=1$ string theory at generic complex momenta, and compare with the matrix model results. As already discussed, the string amplitude, defined by the regularized moduli $z$-integral, is expected to be analytic in the domain 
\ie\label{anad}
{\rm Im}(\omega-\omega_i)>0,~~~i=1,2,3.
\fe
The $z$-integral is manifestly convergent for ${\rm Re}((\omega-\omega_i)^2)<0$. In order to access the ``physical" kinetic regime, i.e. ${\rm Re}(\omega_i)>0$ with $i\epsilon$ prescription obeying (\ref{anad}), we need to move to the domain ${\rm Re}((\omega-\omega_i)^2)>0$ which requires regularization of the $z$-integrand as discussed in section \ref{regz}.

\begin{figure}[h!]
\centering
\includegraphics[width=10cm]{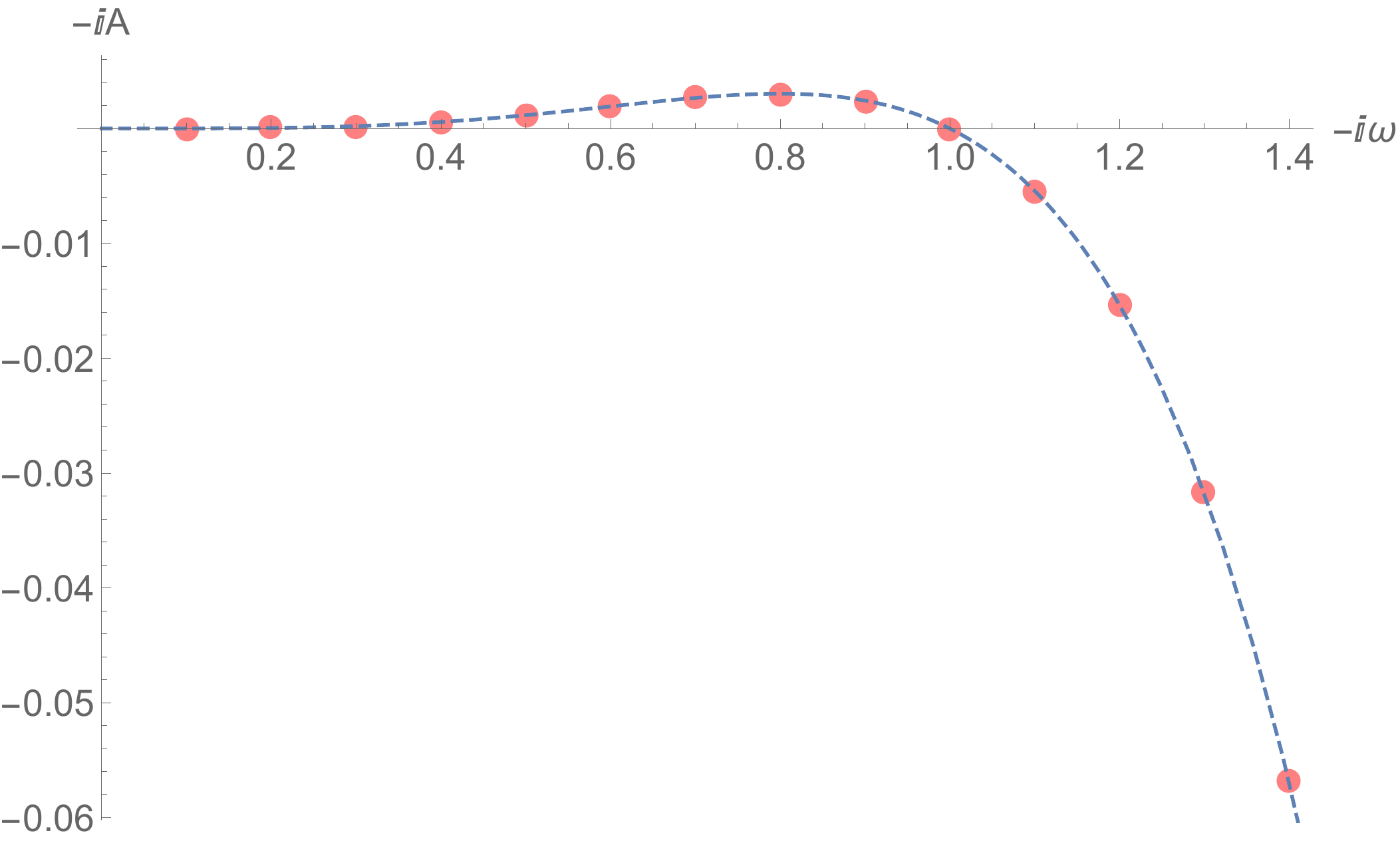}
\caption{Numerical results for the string tree level amplitude $-i{\cal A}_{1\to 3}^{(0)}$ computed for $\omega \in i\mathbb{R}_{>0}$ and $\omega_1=\omega_2=\omega_3 = \omega/3$ (red dots), compared to the matrix model result $-i{\cal A}_{1\to 3}^{(0)}=\omega\omega_1\omega_2\omega_3(1+i\omega)$ (blue dashed line).}
\label{fig:sphere4imw}
\end{figure}

The sphere 4-point Virasoro conformal blocks are computed using Zamolodchikov's recursion relation \cite{Zamolodchikov:1985ie, 1987TMP....73.1088Z}, as an expansion in the elliptic nome $q$, related to the cross ratio $z$ by $q=E(z)$,
\ie\label{enome}
E(x) = \exp\left[-\pi {K(1-x)\over K(x)} \right],~~~{\rm where \,~} K(x)={}_2F_1({1\over 2},{1\over 2}, 1|x).
\fe
The $q$-expansion of the conformal block converges on the entire complex $z$-plane, except for the singular points $z=1$ and $\infty$. The Liouville 4-point functions are then evaluated by numerically integrating the Virasoro conformal blocks with the DOZZ structure constants. This results in a $z$-integrand that is covariant with respect to the crossing transformations $z\to 1-z$ and $z\to 1/z$. It suffices to perform the $z$-integral over the domain $D=\{ |z-1|<1, {\rm Re}(z)<{1\over 2}\}$, and recover the contribution from the rest of the $z$-plane by crossing symmetry. This is useful since the conformal block expansion is in small $q$ (or $z$). For a more detailed discussion see appendix C.2 of \cite{Chang:2014jta}.

To begin with, we consider purely imaginary momenta, with ${\rm Im}\omega>0$ and restrict to the special case $\omega_1=\omega_2=\omega_3 = {\omega/3}$. As we analytically continue the Liouville 4-point function from real momenta, provided that ${\rm Im}(\omega+\omega_i) = {4\over 3}{\rm Im}\omega < 2$, no poles of the structure constants cross the $P$-integration contour and thus (\ref{liouville4pt}) remains valid. The $z$-integral manifestly converges in this case. We truncate the Virasoro conformal blocks up to order 12 in the $q$-expansion, and perform the integration over $P$ and the moduli $(z,\bar z)$ numerically as above. Some further technical details are discussed in Appendix \ref{numericsdetails}. The result, as shown in Figure \ref{fig:sphere4imw}, is in excellent agreement with the matrix model result for ${\cal A}_{1\to 3}^{(0)}$ (\ref{matrixanswer}).

\begin{figure}[h!]
\centering
\subfloat[]{\includegraphics[width=8cm]{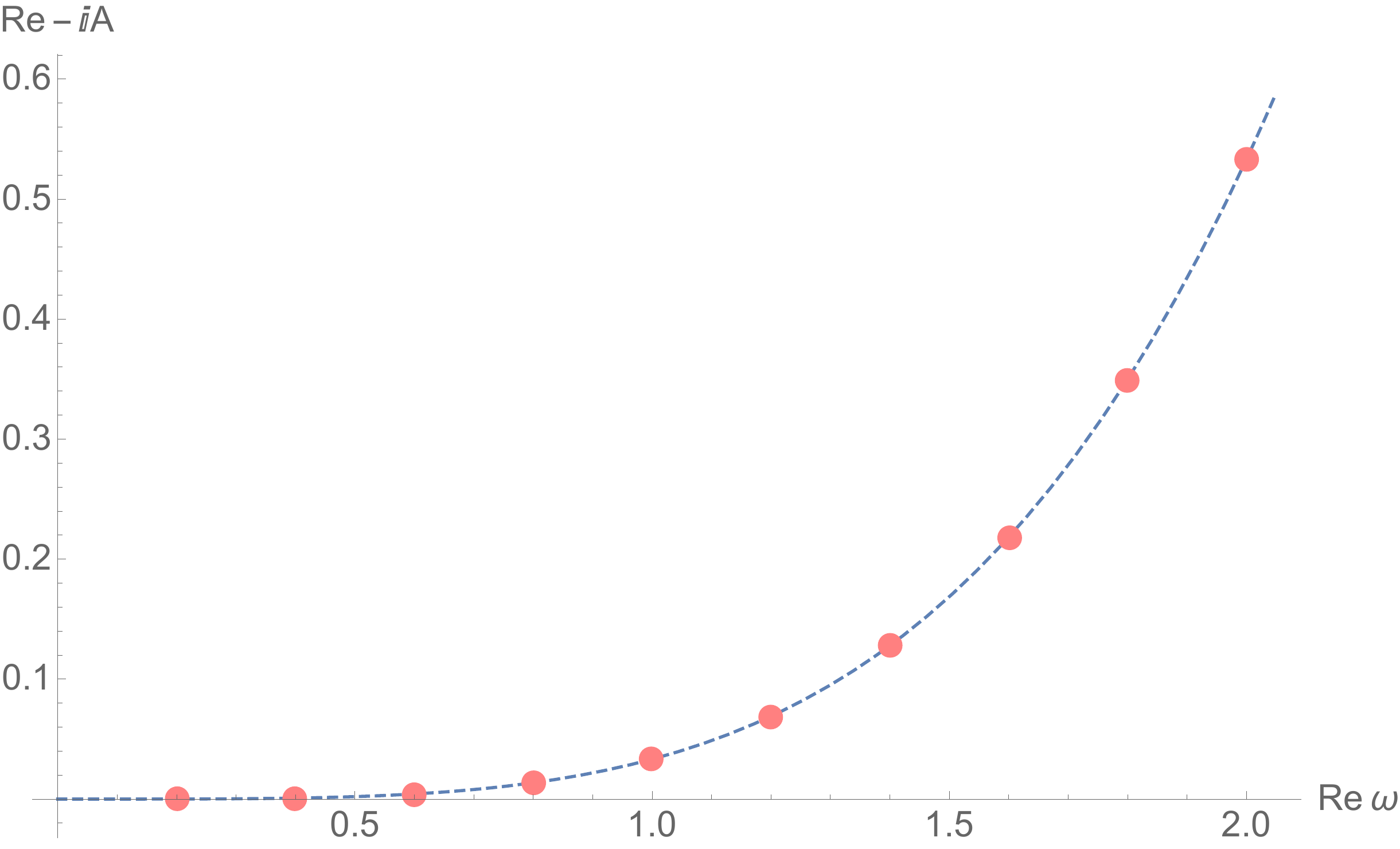}}~~~
\subfloat[]{\includegraphics[width=8cm]{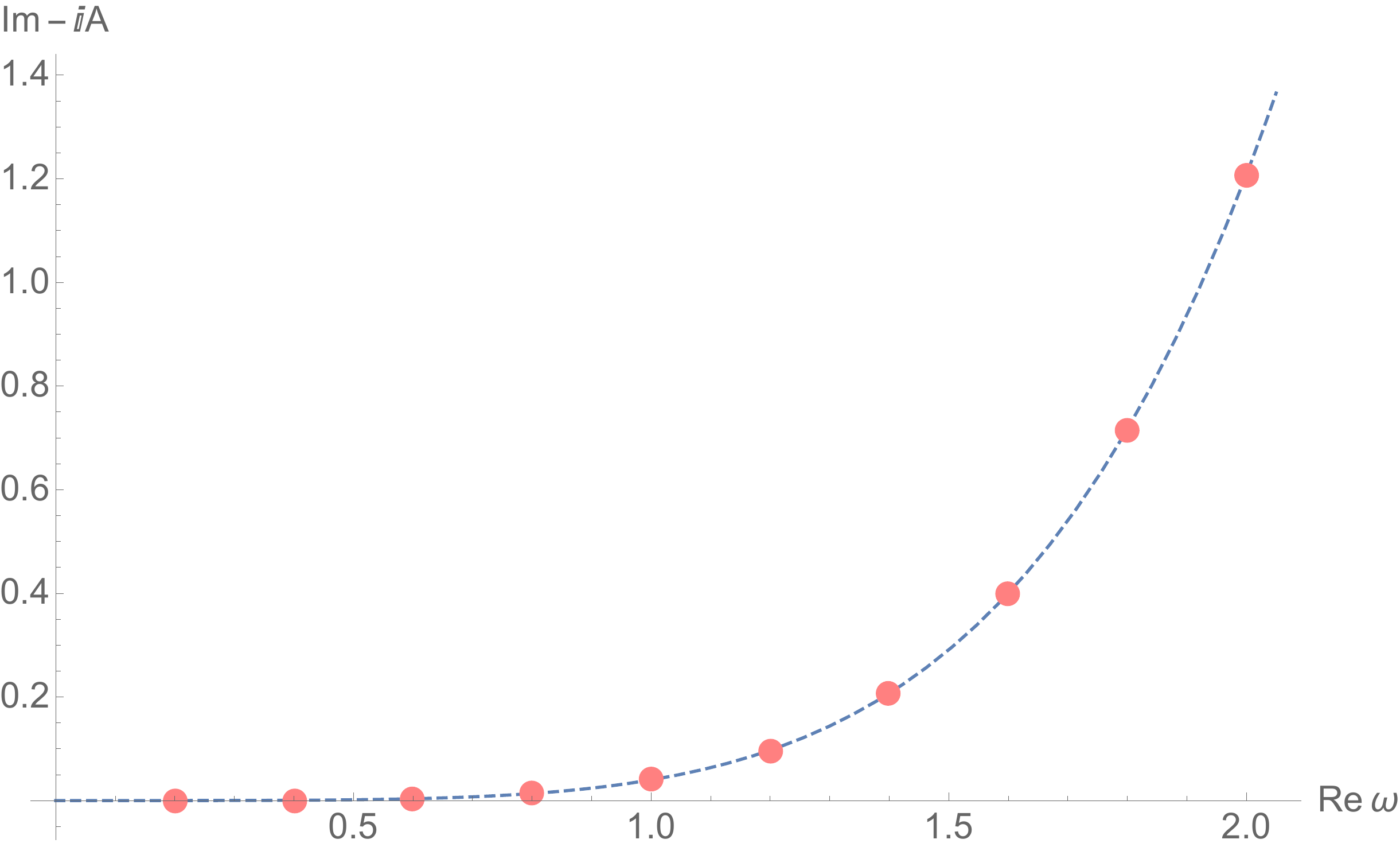}}
\caption{Numerical results for the real and imaginary parts of the string amplitude for $-i{\cal A}^{(0)}_{1\to 3}$ (red dots) in comparison to the matrix model answer (blue dashed line) for $\omega \in \mathbb{R}_{>0}+i\epsilon$ and $\omega_1=\omega_2=\omega_3 = {\omega/3}$, with $\epsilon=0.01$.}
\label{fig:sphere4physw}
\end{figure}

In Figure \ref{fig:sphere4physw}, we present the numerical results for the $1\to 3$ string tree amplitude with $\omega \in \mathbb{R}_{>0}+i\epsilon$, specializing to $\omega_1=\omega_2=\omega_3 = {\omega/3}$, for small positive $\epsilon$. In this case, since ${\rm Re}((\omega-\omega_i)^2)>0$, the regulator (\ref{reg}) is needed to perform the moduli $z$-integral. The result is again in agreement with the matrix model answer (with less than $0.2\%$ discrepancy).

\begin{figure}[h!]
\centering
\subfloat[]{\includegraphics[width=8cm]{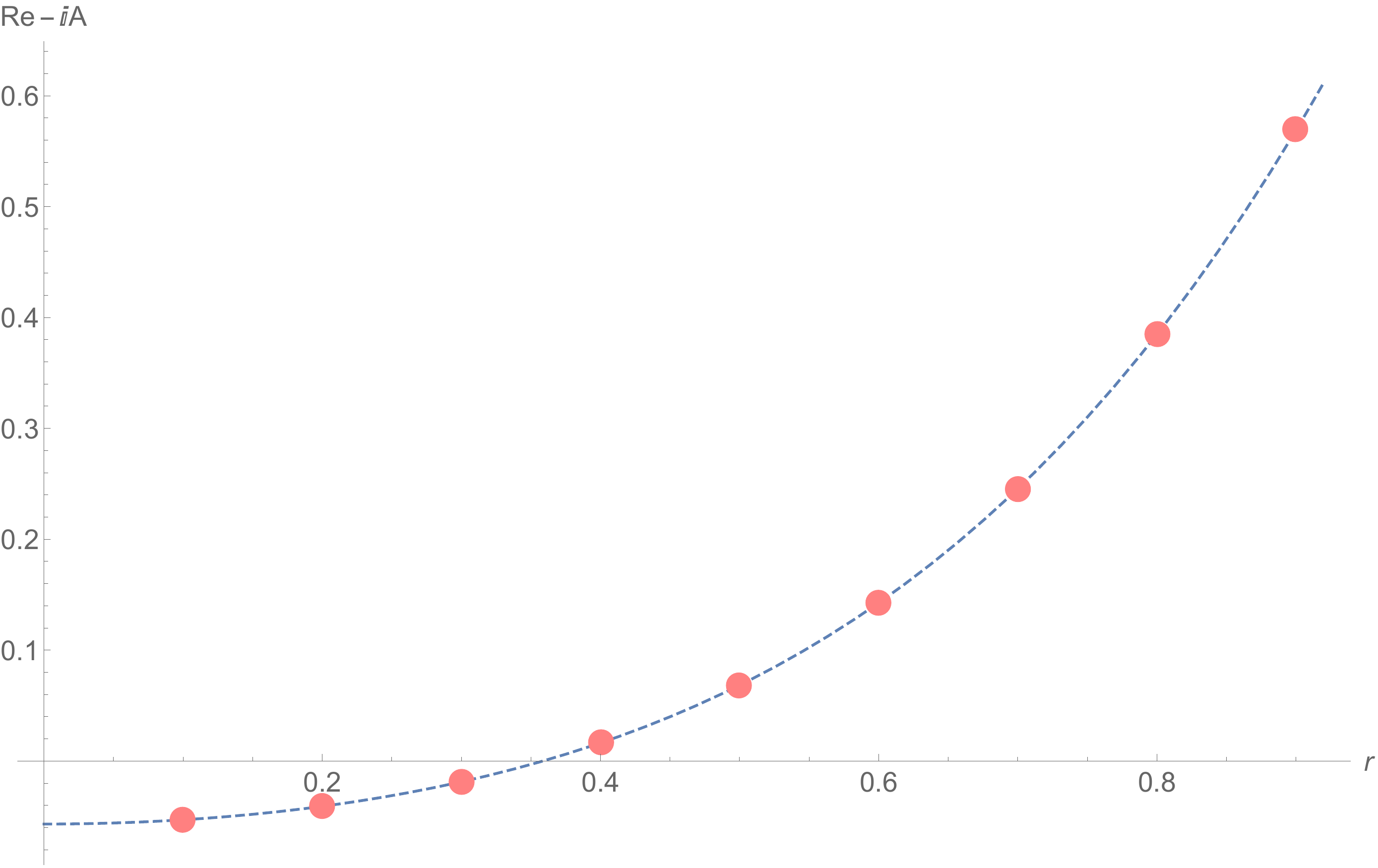}}~~~
\subfloat[]{\includegraphics[width=8cm]{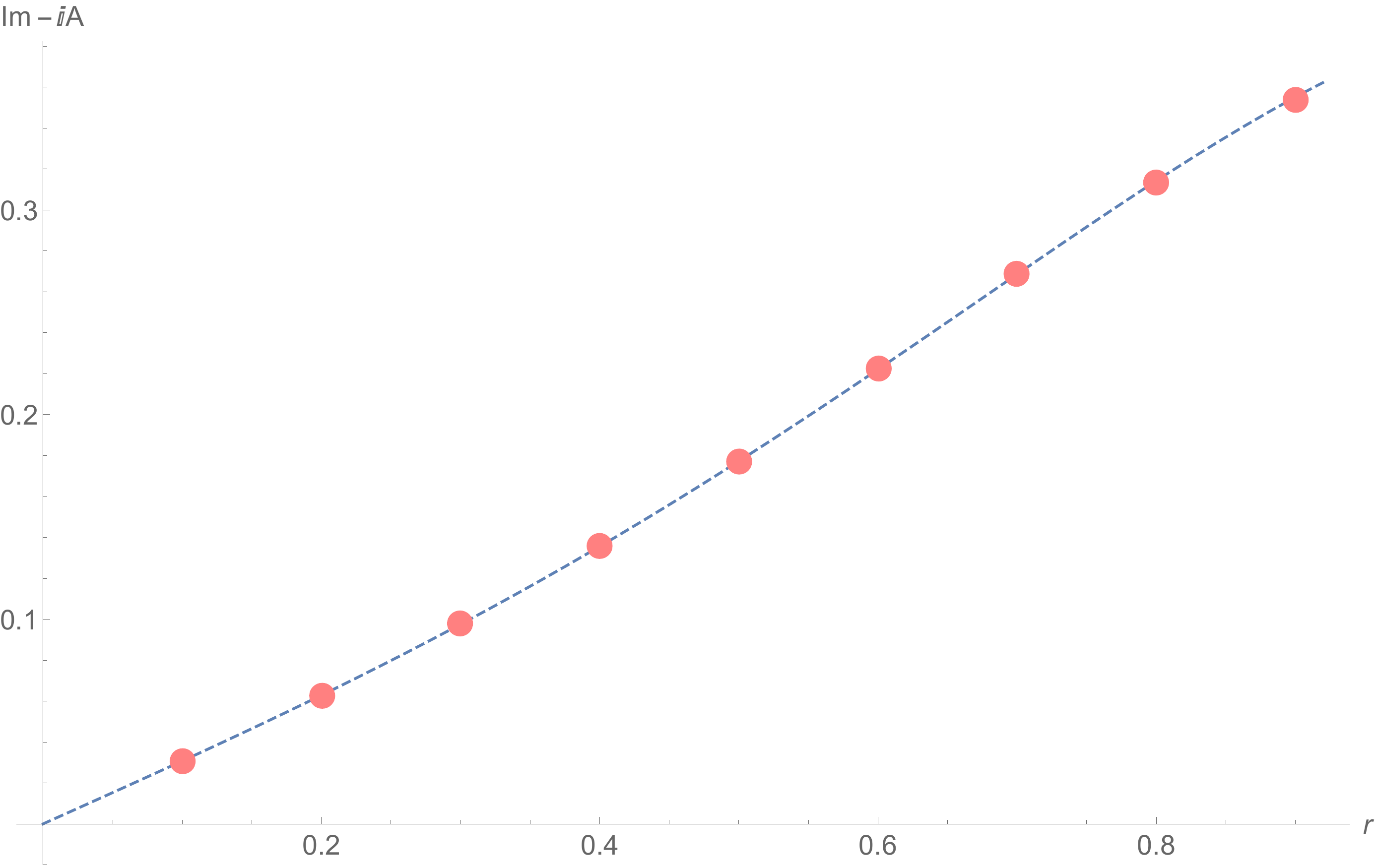}}
\\
\centering
\subfloat[]{\includegraphics[width=8cm]{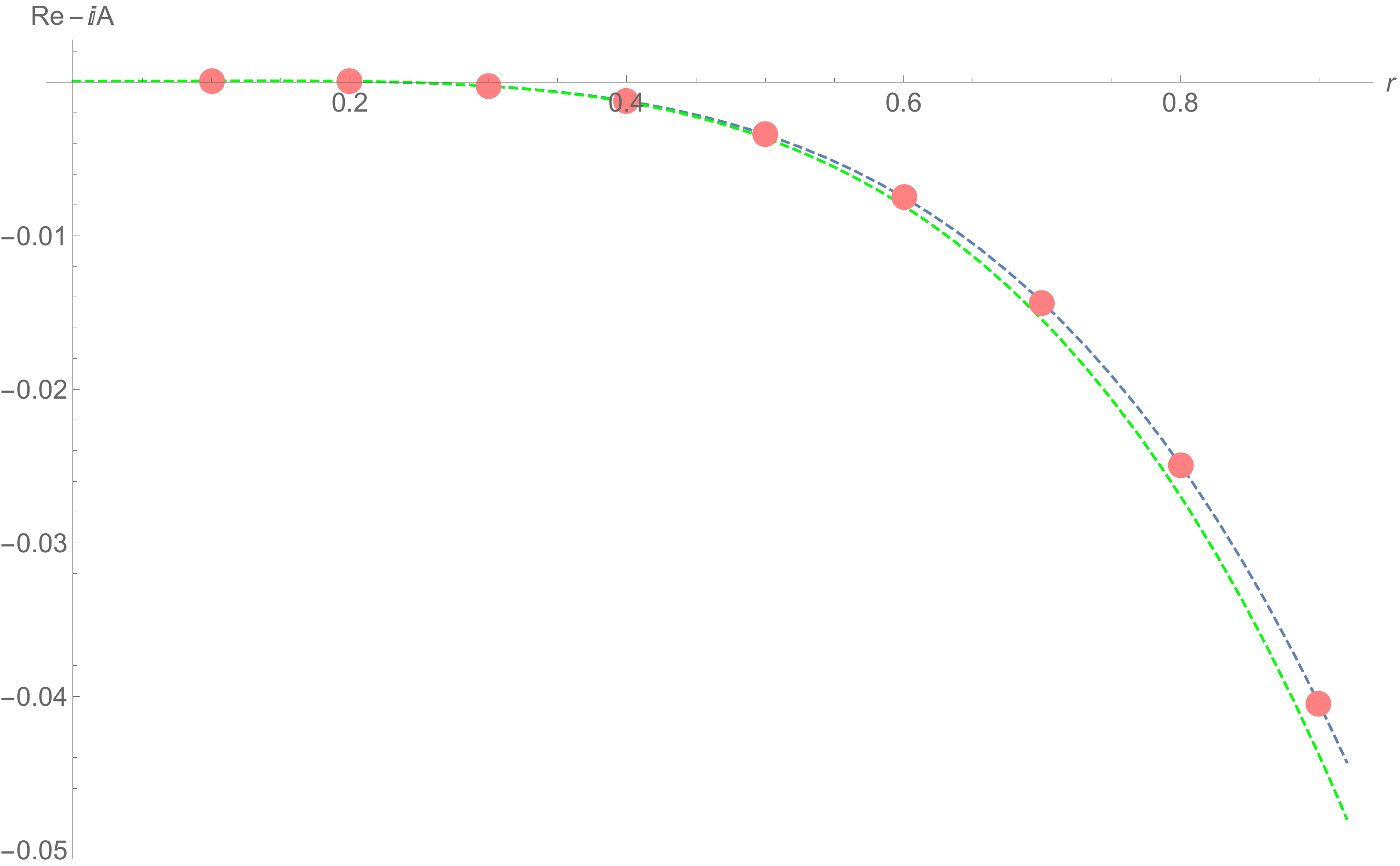}}~~~
\subfloat[]{\includegraphics[width=8cm]{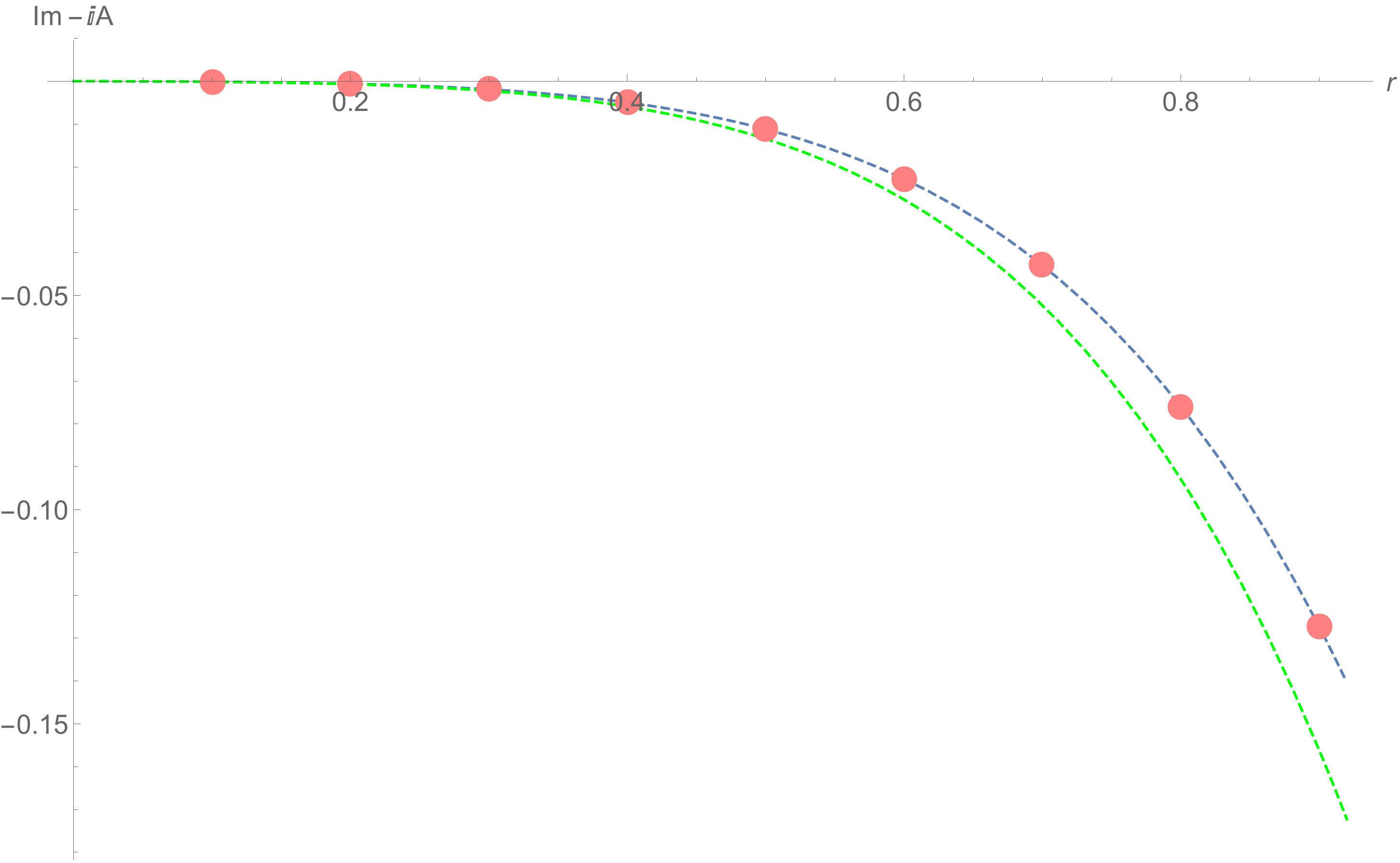}}
\caption{Numerical results for the real and imaginary parts of the string amplitude for $-i{\cal A}^{(0)}_{2\to 2}$ (red dots) in comparison to the anticipated analytic result (\ref{matrix22}) (blue dashed line) at the momenta (\ref{omegachoice}),  with varying $0\leq r\leq 0.9$. The parameter $a$ that controls the imaginary part of the energies is taken to be $a=1.4$ in (a), (b) and $a=0.2$ in (c), (d). The $z$-integral is regularized as in (\ref{reged}) for $r>4a/9$. The green dashed line in (c), (d) corresponds to (\ref{matrix22}) with ${\rm Imax}\{\omega_j\}$ replaced by the $\omega_j$ with the largest real part (namely $\omega_4$ for the momenta (\ref{omegachoice})), which clearly deviates from the string amplitude.}
\label{fig:sphere4genericw}
\end{figure}

Next, we consider $2\to 2$ string tree amplitudes at generic complex momenta $\{\omega_1, \omega_2\} \to \{\omega_3, \omega_4\}$, with the choice\footnote{Note that (\ref{matrix22}) also holds for either signs of ${\rm Im}(\omega_i)$ provided that we define the function Imax($\omega_i$) to pick out the $\omega_i$ with largest $|{\rm Im} (\omega_i)|$.}
\ie\label{omegachoice}
\omega_1=r+ia,~~~\omega_2={r\over 2}-i{a\over 3},~~~\omega_3={r\over 4}+i{a\over 3},~~~\omega_4={5r\over 4}+i{a\over 3},
\fe
for $a=1.4$ and $a=0.2$ respectively, and varying real $r$. This is related to the $1\to 3$ amplitude considered in Figure \ref{fig:sphere4imw} and \ref{fig:sphere4physw} by analytic continuation (where $\omega, \omega_1, \omega_2, \omega_3$ are now relabeled $\omega_1, -\omega_2, \omega_3, \omega_4$). Note that for $r>{4a\over 9}$, the $z$-integral must be regularized according to (\ref{reged}), which maintains analyticity of the amplitude. The numerical results are shown in Figure \ref{fig:sphere4genericw}, which indeed agrees with (\ref{matrix22}). Importantly, ${\rm Imax}\{\omega_j\} = \omega_1$ picks out the complex momenta with the {\it largest imaginary part} in (\ref{omegachoice}), as anticipated from the general analytic structure of the 4-point amplitude discussed earlier.

\section{Genus one 2-point reflection amplitude}

In this section we study the genus one contribution to the $1\to 1$ S-matrix element in $c=1$ string theory. As in the usual bosonic string perturbation theory, this amplitude is given by
\ie\label{torusamp}
& {(2\pi)^2\over 2} \int_{\cal F} d^2\tau \int_{T^2(\tau)} d^2z \left\langle b\tilde b c\tilde c\, {\cal V}_\omega^+(z,\bar z) {\cal V}_{\omega'}^-(0) \right\rangle_{T^2(\tau)} = i{{(2\pi)^2}\over 2} g_s^2 C_{T^2} \delta(\omega-\omega') \int_{\cal F} {d^2\tau\over \sqrt{\tau_2} }|\eta(\tau)|^2
\\
&~~~~~~~~~~~~~~~~~~\times \int_{T^2(\tau)} d^2z \left| {2\pi\over \partial_z\theta_1(0|\tau)} \theta_1\left({z\over 2\pi}|\tau\right) e^{-{({\rm Im} z)^2\over 4\pi \tau_2} } \right|^{\omega^2}
\left\langle V_{\omega\over 2}(z,\bar z) V_{\omega\over 2}(0) \right\rangle_{{\rm Liouville}, T^2(\tau)} .
\fe
Here $C_{T^2}$ is a normalization constant associated with the torus amplitude, which in principle can be fixed in terms of the tree amplitudes via unitarity. We will show in section \ref{sec:torus_res} that in fact $C_{T^2}=1$. ${\cal F}$ is the $PSL(2,\mathbb{Z})$ fundamental domain ${\rm Im}(\tau)>0$, $|{\rm Re}(\tau)|<{1\over 2}$, $|\tau|>1$. The coordinate $z$ on the torus is subject to the identification $z\sim z+2\pi \sim z + 2\pi \tau$. $\left\langle V_{\omega\over 2}(z,\bar z) V_{\omega\over 2}(0)\right\rangle_{{\rm Liouville}, T^2(\tau)}$  is the torus 2-point function in Liouville CFT, that can be computed from the conformal block decomposition in either the OPE channel
\ie{}
&\left\langle V_{\omega\over 2}(z,\bar z) V_{\omega\over 2}(0)\right\rangle_{{\rm Liouville}, T^2(\tau)}
= \int_0^\infty {dP dP'\over\pi^2} {\cal C}({\omega\over 2},{\omega\over 2},P) {\cal C}(P',P',P)
\\
&~~ \times  {\cal F}_{\rm OPE}(1+{\omega^2\over 4}, 1+{\omega^2\over 4}, 1+P^2, 1+P'^2;z,\tau)  \overline{{\cal F}_{\rm OPE}}(1+{\omega^2\over 4}, 1+{\omega^2\over 4}, 1+P^2, 1+P'^2;\bar z,\bar \tau),
\label{liou_ope}
\fe
or the necklace channel
\ie{}
&\left\langle V_{\omega\over 2}(z,\bar z) V_{\omega\over 2}(0)\right\rangle_{{\rm Liouville}, T^2(\tau)}
= \int_0^\infty {dP_1 dP_2\over\pi^2} \left({\cal C}({\omega\over 2},P_1,P_2)\right)^2 
\\
&~~\times {\cal F}_{\rm necklace}(1+{\omega^2\over 4}, 1+{\omega^2\over 4}, 1+P_1^2, 1+P_2^2; z,\tau) \overline{{\cal F}_{\rm necklace}}(1+{\omega^2\over 4}, 1+{\omega^2\over 4}, 1+P_1^2, 1+P_2^2; \bar z,\bar \tau).
\label{liou_neck}
\fe
Here ${\cal F}_{\rm OPE}(d_1, d_2, h, h';z, \tau)$ and ${\cal F}_{\rm necklace}(d_1, d_2, h_1, h_2; z,\tau)$ are the OPE and necklace channel Virasoro conformal blocks in the flat torus frame, as will be reviewed in section \ref{torusblock} below.

\subsection{Limits of the moduli integral}

The moduli integrand as a function of $z$ and $\tau$ is potentially singular in the limits $z\to 0$ or $\tau \to i\infty$. Let us analyze these two limits. For fixed generic $\tau$, the $z$-integral in the vicinity of $z=0$ takes the form
\ie
\int d^2z \int_0^\infty {dP\over\pi} {\cal C}({\omega\over 2},{\omega\over 2},P)\, |z|^{-2+2 P^2} \langle V_P \rangle_{{\rm Liouville}, T^2(\tau)}
\fe
Both ${\cal C}({\omega\over 2},{\omega\over 2},P)$ and the torus 1-point function $\langle V_P \rangle_{{\rm Liouville}, T^2(\tau)}$ have a simple zero at $P=0$. Thus the $P$-integral for small $P$ produces a $z$-integrand of the form $\sim |z|^{-2}(-\log|z|)^{-3/2}$, whose $z$-integral converges near $z=0$.

Let us now inspect the large $\tau_2$ limit, for a generic fixed $z$. In this limit, the torus 2-point function in the Liouville CFT is dominated by an integral of sphere 4-point functions with the insertions of a pair of $V_P$'s with small Liouville momentum $P$. The $\tau_2\to \infty$ limit of the $\tau$-integral takes the form
\ie
\int {d^2\tau\over\sqrt{\tau_2}}  \int_0^{P_*} {dP\over\pi} e^{-4\pi\tau_2 P^2}\left\langle V_{\omega\over 2}(e^{-i z}, e^{i\bar z})  V_{\omega\over 2}(1) V_P(0) V_P(\infty) \right\rangle_{{\rm Liouville}, S^2}
\label{eq:largetau2}
\fe
The 4-point function involved again has a double zero at $P=0$. Thus, the $P$-integral for small $P$ produces a $\tau$-integrand that scales like $\tau_2^{-2}$, giving a convergent $\tau$-integral at large $\tau_2$.

A potentially problematic limit, however, is when $\tau_2$ and ${\rm Im}(z)$ go to infinity simultaneously. Let us write ${\rm Im}(z)\equiv t_1$ and $\tau_2 \equiv t_1 + t_2$. In the regime $t_1, t_2\gg1$, the moduli integral looks like
\ie{}
&\int^\infty {dt_1 dt_2\over \sqrt{t_1+t_2}} \exp\left[ {\pi t_1 t_2\over t_1+t_2}\omega^2 \right]  \int_0^{P_*} {dP_1 dP_2 }\, P_1^2 P_2^2 e^{-4\pi t_1 P_1^2 -4\pi t_2 P_2^2}
\\
& \sim \int^\infty {dt_1 dt_2\over (t_1t_2)^{3/2} \sqrt{t_1+t_2}} \exp\left[ {\pi t_1 t_2\over t_1+t_2}\omega^2 \right].
\fe
We see that this integral is divergent for ${\rm Re}(\omega^2)>0$, and thus a suitable regularization is needed to define (\ref{torusamp}) in the physical momentum regime. Alternatively, we can compute the amplitude by analytic continuation from ${\rm Re}(\omega^2)<0$. In fact, if it weren't for this divergence at real $\omega$, the moduli integral in (\ref{torusamp}) would be real, and could not possibly agree with the matrix model answer for ${\cal A}_{1\to 1}^{(1)}$ in (\ref{matrixanswer}) which has both real and imaginary parts. On the other hand, the matrix model result for ${\cal A}_{1\to 1}^{(1)}$ is purely imaginary when analytically continued to $\omega\in i\mathbb{R}$, which agrees with the moduli integral in (\ref{torusamp}) being real and finite in this case.

\subsection{Torus conformal blocks}\label{torusblock}

An efficient method of computing the torus Virasoro conformal blocks in question is the $c$-recursive representation \cite{Zamolodchikov:1985ie, Hadasz:2009db, Cho:2017oxl}. Rather than focusing on the $c=25$ case, we will consider the analytic continuation of the conformal blocks in $c$. For generic assignments of external and internal weights, the Virasoro conformal block has only simple poles in $c$ with known residues, while the $c\to \infty$ limit is given by the vacuum torus character multiplying the corresponding global $SL(2)$ conformal block. These properties combine to give a set of recursion formulae that relates the Virasoro conformal block of central charge $c$ to ones with an internal weight $h_i$ shifted to $h_i+rs$, for a pair of integers $r\geq 2, s\geq 1$, and at the same time the central charge $c$ replaced by $c_{rs}(h_i)$, the central charge value at which a primary of weight $h_i$ would have a null descendant at level $rs$.

A $c$-recursion formulae for the torus 2-point block in the OPE channel was given in equations (4.35) and (4.36) of \cite{Cho:2017oxl}. Here we simply comment that the formulae of \cite{Cho:2017oxl} computes the OPE channel block as a series expansion in $q=e^{2\pi i \tau}$ and $v = e^{- i z}-1$. For fixed $q$, the $v$ expansion a priori converges in the range $|v|<1-|q|$. For our numerical integration of the torus amplitude, it is more useful to pass to the expansion variable $z$ rather than $v$, as the $z$-expansion converges over the range $|z|<2\pi$ (for $\tau$ in the fundamental domain ${\cal F}$).

An explicit $c$-recursion formula for the torus 2-point block in the necklace channel is given in Appendix \ref{torusblockap}, in the form of an expansion in $q_1 = e^{iz}$ and $q_2 = e^{i(2\pi \tau - z)}$. A priori, the expansion in $q_1$ or $q_2$ has radius of convergence 1. We can extend the convergence range by passing to a new set of expansion variables,
\ie
\widehat q_i = E(q_i),~~~i=1,2,
\fe 
where $E(x)$ is the elliptic nome map as defined in (\ref{enome}).
Now the expansion in $\widehat q_1$ and $\widehat q_2$ converges for $|\widehat q_1|, |\widehat q_2|< |E(q^{-1})|$. If we take $\tau$ to be in the fundamental domain ${\cal F}$, $|E(q^{-1})|> 0.3008$, the $\widehat q_i$-expansion converges in particular for $|z|>0.0187$.

After integrating the internal Liouville momenta to obtain the torus 2-point function, we see that the expansion in $q$ and $z$ through the OPE channel block and the expansion in $\widehat q_1$ and $\widehat q_2$ through the necklace channel block together cover entire moduli space with good convergence property.

\subsection{Torus 2-point amplitude at resonance momenta}
\label{sec:torus_res}

Similarly to the resonance momenta at which the Liouville sphere correlator reduces to that of the linear dilaton CFT, as discussed in section \ref{sec:analytic_tree}, there is a also a set of resonance momenta for the torus correlator and the corresponding genus one string amplitude. We will see that the latter simplifies substantially when the energy of the particle $\omega$ is analytically continued to $\omega=2i$, which allows for an analytic evaluation of the string amplitude at this energy, and fixes the torus normalization constant $C_{T^2}$.

Note that the Liouville vertex operator $V_{\omega\over 2}$ has conformal weight $h=\tilde h = 1+{\omega^2\over 4}$, which vanishes at $\omega\to 2i$. Thus, one may anticipate that the conformal blocks that contribute to the Liouville torus 2-point function reduces to the torus character. We will see that this is indeed the case. Interestingly, the Liouville structure constants vanishes in the $\omega\to 2i$ limit, which compensates for a divergence coming from the moduli integral, resulting in a finite amplitude at $\omega=2i$.

To proceed, let us consider the Liouville torus 2-point function expressed as an integral over conformal blocks in the OPE channel (\ref{liou_ope}), for complex energy $\omega$ with ${\rm Re}(\omega^2)<0$
where the $z$-integral will be manifestly convergent. The structure constant ${\cal C}({\omega\over 2}, {\omega\over 2}, P)$ appearing in (\ref{liou_ope}) has poles in $P$ at
\ie\label{ppoles}
P= ni\pm\omega,~~~ n=\pm 1, \pm 2, \pm 3, \cdots
\fe
The integrand of (\ref{liou_ope}) has other poles in $P$, but they do not play any role in the following. As we analytically continue $\omega$ away from the real axis, some of the poles (\ref{ppoles}) may cross the $P$-integration contour, in which case we must include the residue contributions from such poles to maintain analyticity of the string amplitude (or equivalently, deforming the $P$-contour to avoid the poles). In particular, as we take $\omega$ to $2i$ along the positive imaginary axis, the pole at $P=-i+\omega$ crosses the $P$-contour (had we chosen to deform the $P$-contour instead to avoid the poles, it would be pinched by the pair of poles at $P=-i+\omega$ and at $P=3i-\omega$). The analytically continued Liouville torus 2-point function can thus be written as the sum of the residue contribution at $P=-i+\omega$ together with the original $P$-contour integral as in (\ref{liou_ope}). 

Writing $\omega=2i(1-\epsilon)$ and taking the $\epsilon\to 0^+$ limit, we find that the contribution to the string amplitude from the integral over the original real $P$ contour vanishes. This is due to the fact that ${\cal C}({\omega\over 2},{\omega\over 2},P)$ vanishes like $\epsilon^4$ for $P>0$, and after the $P$-integral the Liouville correlator vanishes like $\epsilon^3$. The $z$-integral, as we will see shortly, introduces a divergent factor $\sim \epsilon^{-1}$, but the contribution to the string amplitude still vanishes like $\epsilon^2$.

This leaves the residue at $P=-i+\omega = i(1-2\epsilon)$ as the only contribution to the string amplitude in the $\omega\to 2i$ limit, which we now compute. The Liouville correlator in the $\epsilon\to 0^+$ limit reduces to
\ie \label{liouvillelimit}
&\left\langle V_{i(1-\epsilon)}(z,\bar z) V_{i(1-\epsilon)}(0)\right\rangle_{{\rm Liouville}, T^2(\tau)}
\\
&\to (-2i)\lim_{\epsilon\to 0}\int_0^\infty \frac{dP'}{\pi} {\rm Res}_{P=i(1-2\epsilon)} 
{\cal C}(i(1-\epsilon), i(1-\epsilon), P) {\cal C}(P',P',P) 
\\
&~~~~\times {\cal F}_{\rm OPE}(2\epsilon-\epsilon^2,2\epsilon-\epsilon^2, 1+P^2, 1+P'^2;z,\tau) \overline{{\cal F}_{\rm OPE}}(2\epsilon-\epsilon^2,2\epsilon-\epsilon^2, 1+P^2, 1+P'^2;\bar z,\bar\tau)
\\
&\to -32\epsilon\int_0^\infty \frac{dP'}{\pi} \frac{|q|^{2P'^2}}{|\eta(\tau)|^2}=-\frac{8\epsilon}{\pi\sqrt{\tau_2}|\eta(\tau)|^2},
\fe
where we used the property that the zero external weight and $P\to i$ limit of the torus 2-point conformal block reduces to the torus Virasoro character of a primary of weight $h=\tilde h = 1+P'^2$ and central charge $c=25$.

The only contribution to the string amplitude that survives the $\epsilon \to 0^+$ limit comes from the integral over $z$ near the origin, where the torus 2-point function from the time-like free boson $X^0$ contributes a factor
\ie{}
&\int d^2z \left|\frac{2\pi}{\partial_z\theta_1(0|\tau)}\theta_1\left(\frac{z}{2\pi}|\tau\right){\rm exp}\left[ -\frac{({\rm Im} z)^2}{4\pi\tau_2}\right]\right|^{-4(1-2\epsilon)^2} 
\\
& \to  \int d^2z  \frac{|z|^{-4+8\epsilon}({\rm Im}z)^2}{\pi\tau_2} \sim {1\over 8 \epsilon \tau_2 }.
\fe
Putting this together with (\ref{liouvillelimit}) in (\ref{torusamp}), we find a finite $\epsilon\to 0^+$ limit, giving
\ie
g^2{\cal A}^{(1)}_{1\to1}\big|_{\omega=2i}=-2\pi iC_{T^2}g_s^2 \int_{\cal F} \frac{d^2\tau}{\tau_2^2}=-\frac{2\pi^2}{3}iC_{T^2}g_s^2,
\fe
where $g_s$ is related to $g$ via (\ref{eq:norms1}). If we assume this matches with the matrix model result ${\cal A}_{1\to 1}^{(1)} = {1\over 24} \left( i \omega^2 + 2i \omega^4 - \omega^5 \right)=-{i\over6}$ at $\omega=2i$, we would fix the normalization constant
\ie
C_{T^2}=1.
\label{torusnorm}
\fe
As remarked in the introduction, the $\omega^5$ term in ${\cal A}_{1\to 1}^{(1)}$ is determined by the tree amplitude ${\cal A}_{1\to 2}^{(0)}$ through unitarity. In the next section, we will verify numerically for generic imaginary $\omega$ that indeed the full $\omega$-dependence of ${\cal A}_{1\to 1}^{(1)}$ is reproduced from the genus one string amplitude.

\subsection{Numerical evaluation at generic imaginary momenta}

We now evaluate numerically the genus one $1\to 1$ amplitude in the domain ${\rm Re}(\omega^2)<0$, where the moduli integral is manifestly convergent and no regularization is required. In order to maintain analyticity in the momenta, we must ensure that while we are taking the external Liouville momenta (${\omega\over 2}$ in this case) to be complex, no poles in the structure constants cross the integration contour in the internal Liouville momenta. This is ensured in both the OPE and necklace channels provided that $|{\rm Im}\,\omega|<1$.

The numerical integration is performed by sampling uniformly in the fundamental domain for $(\tau, \bar\tau)$ with respect to the measure $\int d^2\tau/\tau_2^2$, and for each sampling value of $(\tau, \bar\tau)$ we perform the $z$-integral over half of the torus, ${\rm Re}(z)\in [0,2\pi]$, ${\rm Im}(z)\in [0,\pi\tau_2]$ (note that $\tau_2>{\sqrt{3}\over 2}$). For a small positive number $\epsilon$, we evaluate the torus 2-point function within the disc $|z|<2\pi \epsilon$ by integrating the OPE channel conformal block, and outside the disc by integrating the necklace channel block. Effectively, the expansion parameter in the OPE channel is $\epsilon$, while the expansion parameter in the necklace channel (where the conformal block is computed as a series in the $\hat q_i$ variables) is $|E(e^{2\pi i\epsilon})/E(q^{-1})|$, with radius of convergence 1 in both cases. For instance, with the choice $\epsilon=0.2$, $|E(e^{2\pi i\epsilon})/E(q^{-1})|$ is bounded from above by $|E(e^{2\pi i\epsilon})/E(e^{2\pi e^{\pi i /6}})|\approx 0.1996$, and we can expect to achieve $\sim 10^{-5}$ accuracy by going to level 8 in both channels. In the numerical computation below, the choice $\epsilon=0.15$ is adopted (we have numerically verified the agreement of the OPE channel versus the necklace channel computation of the Liouville 2-point function for $|z|=2\pi \epsilon$ around this value of $\epsilon$).

\begin{figure}[h]
\centering
\includegraphics[width=12cm]{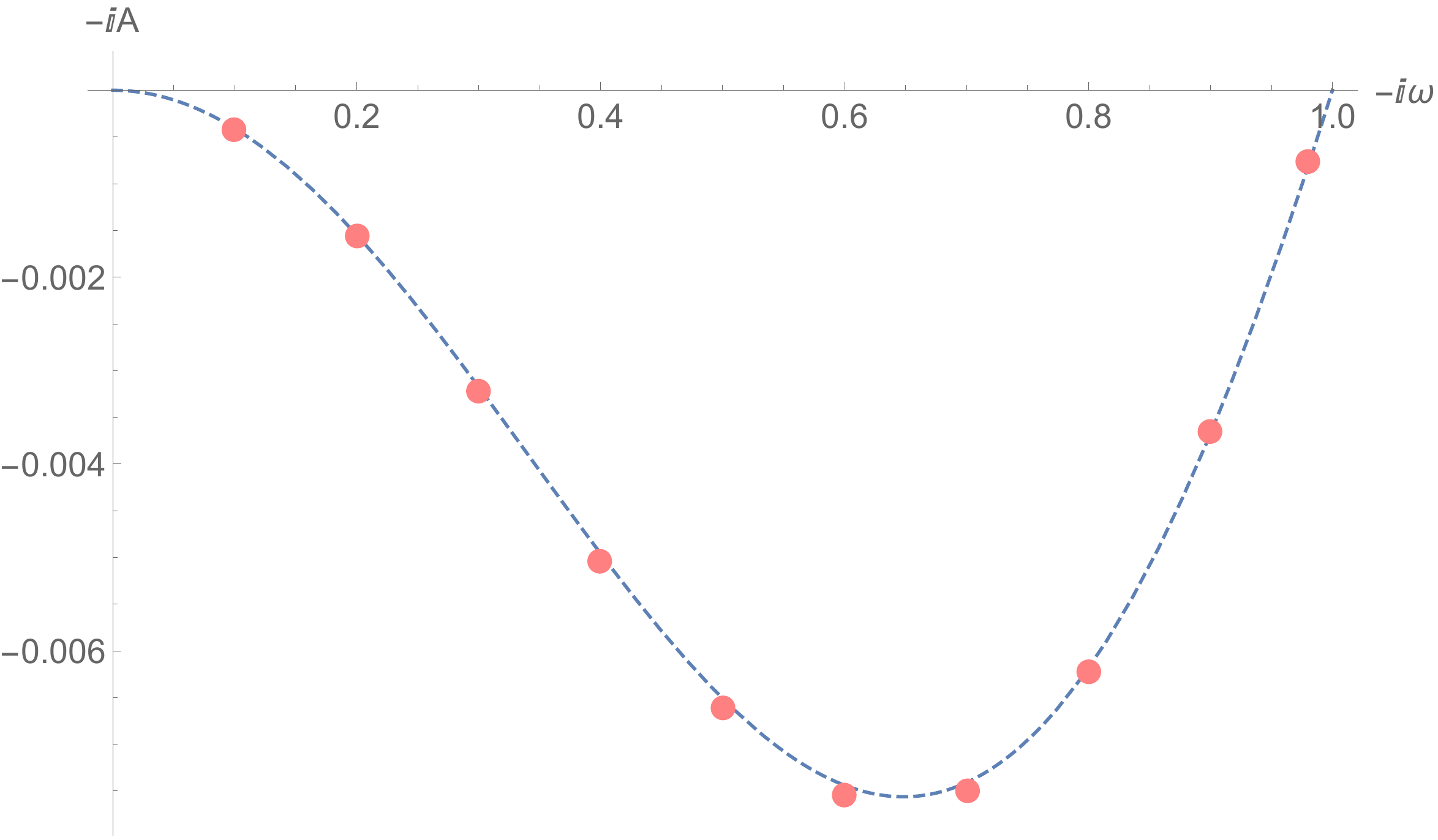}
\caption{Numerical results for genus one 2-point string amplitude $-i{\cal A}_{1\to 1}^{(1)}$ with purely imaginary $\omega = i x$, $x>0$ (red dots) versus the matrix model result $-i{\cal A}_{1\to 1}^{(1)} = {1\over 24}(-x^2+2x^4-x^5)$ (blue dashed line). }
\label{fig:torus4imw}
\end{figure}

To proceed, we take $\omega$ to be purely imaginary, and evaluate the Liouville torus 2-point function as an integral of the relevant Virasoro conformal blocks truncated in their $q$-expansions, over a pair of internal Liouville momenta, and then numerically integrate $z$ over the torus, and finally integrate the torus modulus $\tau$ over the fundamental domain. It suffices to perform the $z$-integral over the ``lower half" of the torus, $0<{\rm Im} z<\pi \tau_2$, where, for $\tau$ taking value in its fundamental domain, good numerical precision is achieved by merely keeping up to level 1 terms in the $\hat{q}_2$ expansion and level 4 in the $\hat{q}_1$ expansion for the necklace channel conformal block, and level 1 in the $q$ expansion and level 4 in $z$ expansion in the OPE channel conformal block (for $|z|<2\pi \epsilon$). 

Our result is shown in Figure \ref{fig:torus4imw}. The numerical result for ${\cal A}_{1\to 1}^{(1)}$ fits a function of the form ${1\over 24}(a i\omega^2 + 2 b i\omega^4 - c \omega^5)$ with $a=1.018$, $b=1.028$, $c=1.0344$. This is in reasonably good agreement with the matrix model result (\ref{matrixanswer}) which corresponds to $a=b=c=1$. The small discrepancy (up to $\sim 2\%$ in the amplitude) is presumably due to numerical errors in the evaluation of the above described 6-fold integral, where a number of interpolations are employed for numerical efficiency. Further details of the numerics are given in Appendix \ref{numericsdetails}.

\section{Discussion}

In this work, we clarified a number of issues regarding the S-matrix in $c=1$ string theory from the worldsheet perspective. The so-called ``leg factors" \cite{Klebanov:1991qa, Ginsparg:1993is} are automatically taken into account once the Liouville vertex operators are properly delta-function normalized\footnote{ 
The leg-pole factor still appears in the asymptotic wave functional of the Liouville vertex operator. An interesting spacetime gravitational interpretation has been given to these Liouville phase factors in \cite{Natsuume:1994sp}, but the computations considered in this paper have nothing to add with regard to this point.}, and the (non-)analyticity of the perturbative string amplitudes are elucidated through the examples of the tree level 4-point amplitude as well as the genus one 2-point amplitude.
The general S-matrix elements of c = 1 string theory is plagued by ambiguities associated with massless scattering in 1+1 dimensions, but such ambiguities can be tamed by working with complex momenta and carefully taking the real momentum limit. As we have seen, the corresponding discontinuities in the S-matrix elements are necessary consequences of unitarity.

We performed direct numerical integration of the Liouville correlators to obtain the tree level $1\to 3$ string amplitude and the genus one $1\to 1$ string amplitude for generic complex momenta in a suitable range (primarily for convenience in regularizing the moduli integral). In the former case, excellent agreement with the matrix model result was found, confirming results previously obtained by analytic continuation from resonance momenta \cite{DiFrancesco:1991ocm, DiFrancesco:1991daf}. In the latter case of the genus one amplitude, a rather difficult 6-fold integration over a pair of internal Liouville momenta and 4 real moduli is performed, resulting in a reasonably good agreement with the matrix model answer up to $\sim 2\%$ error. We find this a rather convincing piece of evidence that the equivalence of the S-matrices on the two sides of the duality extend beyond tree level.

The simplicity of the matrix model results suggests that an analytic derivation of the string amplitudes for generic momenta should be possible, and it is somewhat unsatisfying that thus far we can only compute the latter numerically except at special resonance momenta. An analytic approach may be possible along the lines of \cite{Zamolodchikov:2003yb, Belavin:2005yj, Belavin:2006ex}. We also wish to extend such computations to beyond genus one, where relatively efficient methods of evaluating Virasoro conformal blocks (and hopefully Liouville correlators) on higher genus Riemann surfaces are now available \cite{Cho:2017oxl}.

One expects that any large $N$ gauged matrix model should be dual to some sort of string theory \cite{Aharony:1999ti}. A simple class of models is the $U(N)$ gauged Hermitian one-matrix quantum mechanics, which may be described as a fermi droplet in a 2-dimensional phase space. The $c=1$ matrix model is a special case, which may be viewed as the description of a small part of the phase space near a point on the fermi surface of a very large droplet. One may consider an infinite parameter family of deformations of the fermi surface as well as deformations of the matrix model Hamiltonian. These deformations are expected to be dual to marginal deformations on the worldsheet of the $c=1$ string theory, either by delta-function normalizable vertex operators which corresponds to deforming the background fermi surface in a time-dependent manner \cite{Moore:1992gb}, or by non-normalizable ``special states" \cite{Moore:1991zv, Witten:1991zd} that correspond to deformations of the matrix model Hamiltonian. Presumably, such deformations can be made exactly marginal and would lead to deformed worldsheet CFTs in which the time-like free boson and the $c=25$ Liouville theory are coupled in a nontrivial way. Much less is known about these deformed CFTs; the only established example we are aware of is sine-Liouville theory or its T-dual $SL(2,\mathbb{R})/U(1)$ ``cigar" CFT \cite{Hori:2001ax, Hikida:2008pe, Witten:1991yr, Mandal:1991tz}. We hope that progress in the conformal bootstrap of 2D irrational CFTs will lead to a more precise understanding of the correspondence between general Hermitian one-matrix models and deformed two-dimensional non-critical string theories.

\section*{Acknowledgements}

We would like to thank Scott Collier, Igor Klebanov, Juan Maldacena, Joe Polchinski, Nati Seiberg, and Andy Strominger for discussions and/or comments on a preliminary draft. We are especially grateful to Minjae Cho for sharing his Mathematica code for the numerical evaluation of torus correlation functions in Liouville CFT, and to Ying-Hsuan Lin for sharing his Mathematica code implementing Zamolodchikov's recursion relations. XY thanks Simons Collaboration Workshop on Numerical Bootstrap at Princeton University, ``Quantum Gravity and the Bootstrap" conference at Johns Hopkins University, Perimeter Institute, and Kavli Institute for Theoretical Physics for their hospitality during the course of this work. This work is supported by a Simons Investigator Award from the Simons Foundation and by DOE grant DE-FG02-91ER40654. VR is supported by the National Science Foundation Graduate Research Fellowship under Grant No. DGE1144152. The numerical computations in this work are performed on the Odyssey cluster supported by the FAS Division of Science, Research Computing Group at Harvard University.

\appendix

\section{Torus 2-point conformal blocks}\label{torusblockap}

In this Appendix we give the explicit $c$-recursion for torus 2-point Virasoro conformal block in the necklace channel, following the general prescription of \cite{Cho:2017oxl}. We write $c=1+6(b+b^{-1})^2$, define
\ie
A_{rs}^c = {1\over 2} \prod_{m = 1-r}^r \prod_{n=1-s}^s (mb + n b^{-1})^{-1},~~~ (m,n)\not=(0,0),\, (r,s),
\fe
as well as the fusion polynomials
\ie
P_c^{rs} \begin{bmatrix}d_1\\d_2\end{bmatrix} = \prod_{p=1-r~{\rm step}~2}^{r-1} \prod_{q=1-s~{\rm step}~2}^{s-1}
{\lambda_1+\lambda_2 + pb + q b^{-1}\over 2} {\lambda_1-\lambda_2 + pb + q b^{-1}\over 2},
\fe
where $\lambda_i$ are related to the weights $d_i$ by $d_i = {1\over 4}(b+b^{-1})^2 - {1\over 4}\lambda_i^2$. For given weight $h$, the central charge at which the weight $h$ representation has a level $rs$ null state is
\ie
c_{rs}(h) = 1+6(b_{rs}(h) + b_{rs}(h)^{-1})^2, ~~~ {\rm with}~ b_{rs}(h)^2 = {rs - 1 + 2h+ \sqrt{(r-s)^2 + 4(rs-1) h + 4h^2}\over 1-r^2},
\fe
where $r\geq 2$ and $s\geq 1$.

The necklace block $F$ is a function of the central charge $c$, external weights $d_1, d_2$, internal weights $h_1, h_2$, and the two cylinder moduli parameters $q_1=e^{iz}$, $q_2 = e^{i(2\pi\tau-z)}$. It obeys the recursion relation
\ie
F &= U_c + \sum_{r\geq 2, s\geq 1} \left[ - {\partial c_{rs}(h_1)\over \partial h_1} \right] {q_1^{rs} A_{rs}^{c_{rs}(h_1)} P^{rs}_{c_{rs}(h_1)}\begin{bmatrix} h_2 \\ d_1 \end{bmatrix} P^{rs}_{c_{rs}(h_1)}\begin{bmatrix}h_2\\ d_2 \end{bmatrix}  \over c- c_{rs}(h_1)} F(h_1\to h_1+rs, c\to c_{rs}(h_1))
\\
&~~~ + \sum_{r\geq 2, s\geq 1} \left[ - {\partial c_{rs}(h_2)\over \partial h_2} \right] {q_2^{rs} A_{rs}^{c_{rs}(h_2)} P^{rs}_{c_{rs}(h_2)}\begin{bmatrix}h_1\\ d_1 \end{bmatrix} P^{rs}_{c_{rs}(h_2)}\begin{bmatrix}h_1\\ d_2 \end{bmatrix}  \over c- c_{rs}(h_2)} F(h_2\to h_2+rs, c\to c_{rs}(h_2)),
\fe
from which a series expansion of $F$ in $q_1$ and $q_2$ can be extracted efficiently.
The regular term $U_c$ is given by the product of torus vacuum character with the global $SL(2)$ necklace block,
\ie
U_c = \left[ \prod_{n=2}^\infty {1\over 1-q^n} \right] \sum_{j,k=0}^\infty q_1^j q_2^k { s_{jk}(h_1, d_1, h_2) s_{kj}(h_2, d_2, h_1)\over j! k! (2h_1)_j (2h_2)_k}.
\fe
Here we have defined 
\ie
s_{jk} (h_1, h_2, h_3) = \sum_{p=0}^{{\rm min}\{j,k\}} {j!\over p!(j-p)!} (2h_3+k-1)^{(p)} k^{(p)} (h_3+h_2-h_1)_{k-p} (h_1+h_2-h_3+p-k)_{j-p},
\fe
where $(a)^{(p)} \equiv a(a-1)\cdots (a-p+1)$ is the descending Pochhammer symbol, and $(a)_n\equiv a(a+1)\cdots (a+n-1)$ is the ascending Pochhammer symbol.

\section{Further details of the numerics}
\label{numericsdetails}

\subsection{On the sphere 4-point amplitude}

In numerically evaluating the string tree level 4-point amplitude (\ref{amp4pt}), one must be careful with integration near $z=0$ (or $z=1,\infty$ related by crossing), where the integrand oscillates rapidly. In practice we deal with this by cutting out a small region $D_\delta=\{|z-1|<1,\,0<{\rm Re}(z)<\delta\}$ inside the integration domain $D=\{|z-1|<1,\,0<{\rm Re}(z)<1/2\}$ (as described earlier, the rest of the integral over the complex $z$-plane can be recovered by crossing symmetry). Outside $D_\delta$ one can numerically integrate reliably. The contribution from the region $D_\delta$ can be obtained by expanding the integrand to leading order in $\delta$ and performing the integral analytically. 

\begin{figure}
\centering
\subfloat[]{\includegraphics[width=8cm]{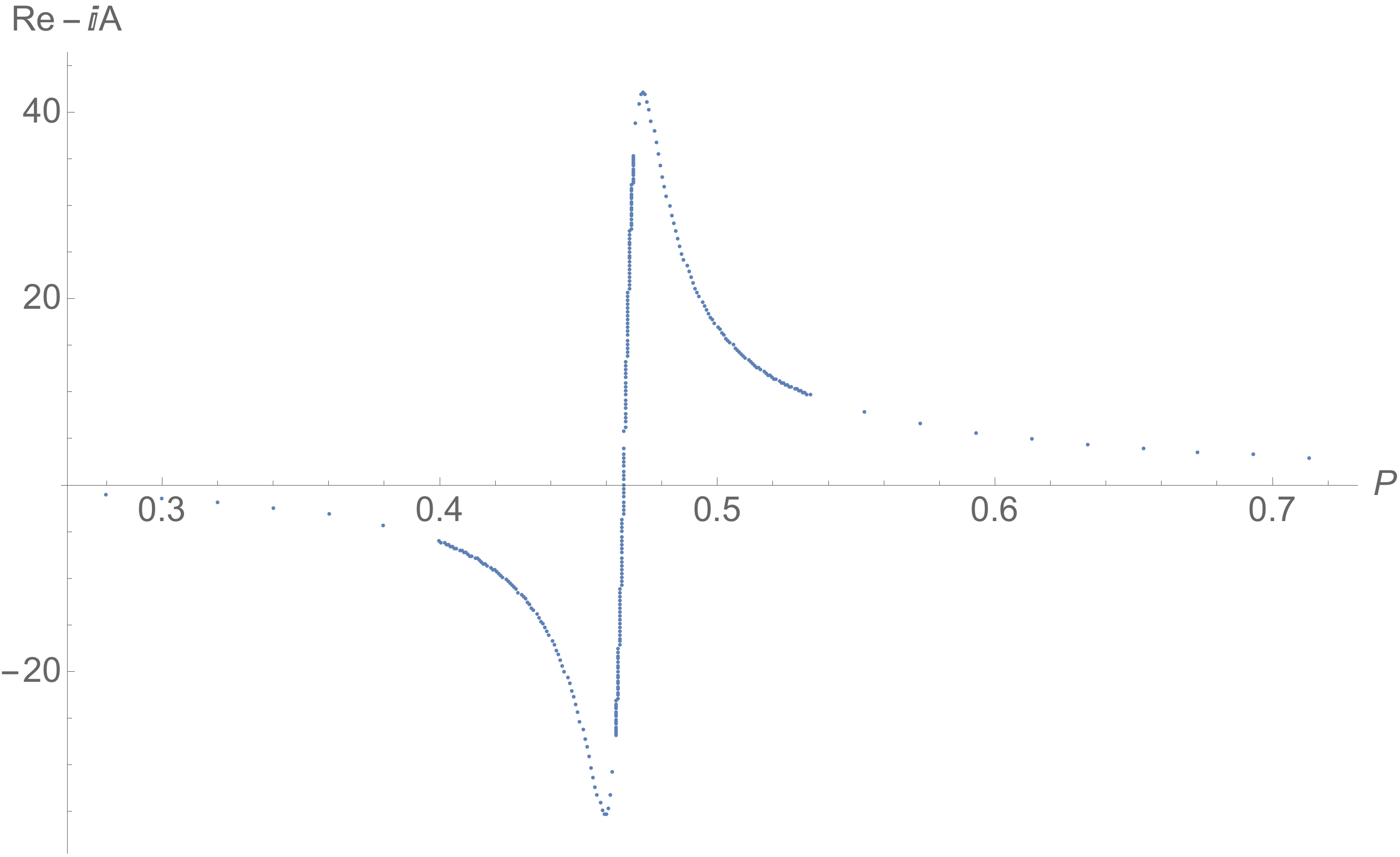}}
\subfloat[]{\includegraphics[width=8cm]{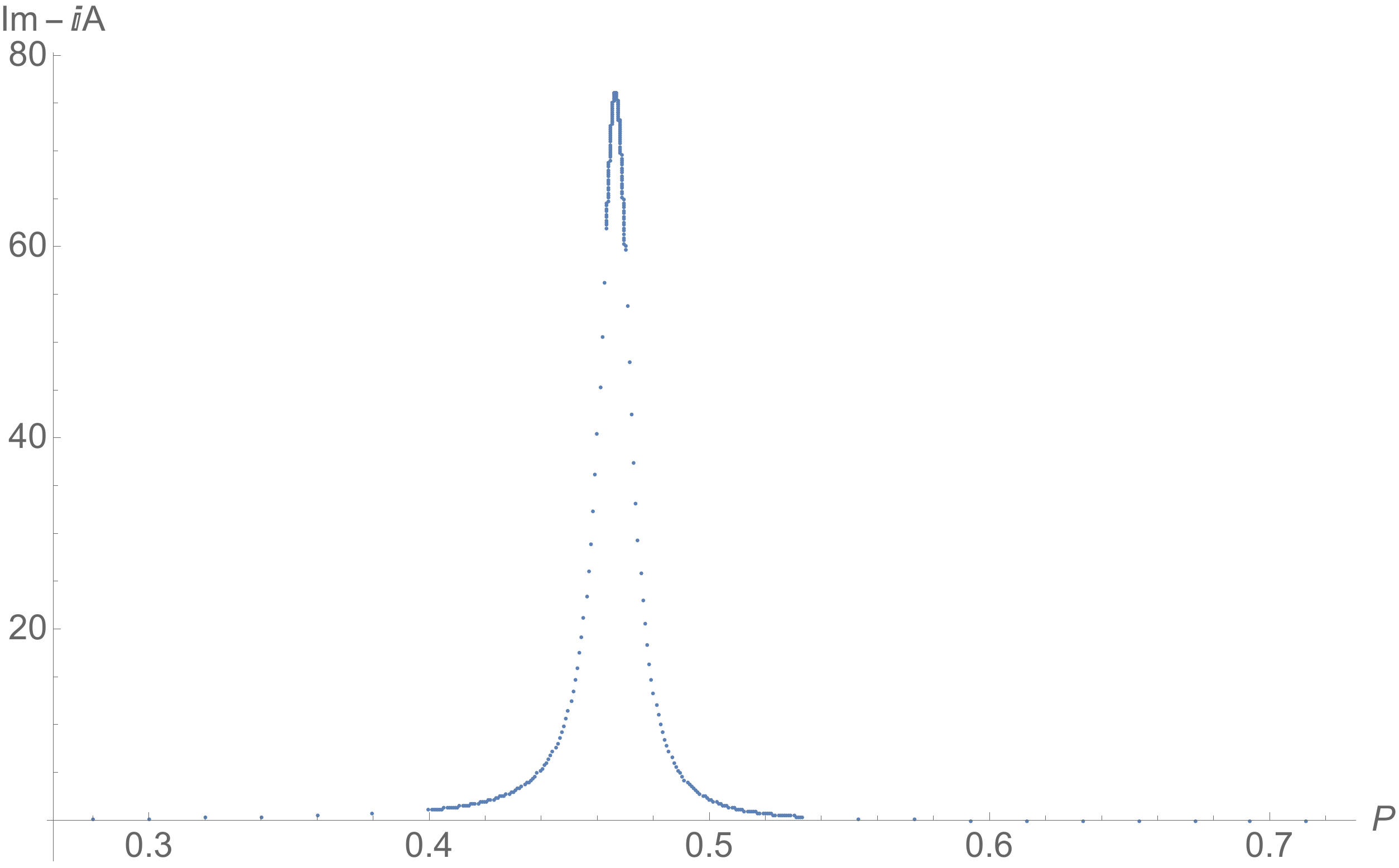}}
\caption{Contributions to the real (left) and imaginary (right) parts of the amplitude $-i{\cal A}_{1\to 3}^{(0)}$ from a range of Liouville momentum $P$, after having already performed the $z$-integral in the domain $D=\{|z-1|<1,\,0<{\rm Re}(z)<1/2\}$. In this example we take $\omega=1.4+i\epsilon$, $\omega_1=\omega_2=\omega_3=\omega/3$, with $\epsilon=0.01$, and the Virasoro conformal block was computed up to order 12 in its $q$ expansion. }
\label{fig:sphere4sample}
\end{figure}

For complex momenta in the regime ${\rm Re}((\omega-\omega_i)^2)>0$, we must regularize the $z$-integral as prescribed in section \ref{regz}. The contribution from $D_\delta$ becomes increasingly important near $P={1\over 2}\sqrt{{\rm Re}((\omega-\omega_i)^2)}$, and in fact will account for most of the imaginary part of the amplitude $-i {\cal A}_{1\to 3}^{(0)} = \omega\omega_1\omega_2\omega_3(1+i\omega)$ when the momenta approach the real axis. As we approach the ``physical regime" where the momenta are close to being real, we must take care of the $P$-integral near $P={1\over 2}\sqrt{{\rm Re}((\omega-\omega_i)^2)}$ by sampling sufficiently finely. An example is shown in Figure \ref{fig:sphere4sample}.

\subsection{On the torus 2-point amplitude}

The computation of the genus one 2-point string amplitude (\ref{torusamp}) involves a rather daunting looking 6-fold integral. The integration over a pair of internal Liouville momenta poses no difficulty as the integrand (Liouville structure constants multiplied by the torus 2-point conformal block) is a smooth function in the momenta. It suffices to evaluate the DOZZ coefficients once for a fixed external energy $\omega$ and a sufficiently fine set of internal momenta. 

We can perform a number of nontrivial consistency checks on the evaluation of Liouville torus 2-point function, including the agreement of the integrals of OPE channel versus the necklace channel conformal blocks over internal Liouville momenta, and the modular covariance of the resulting torus correlator, namely
\ie
& \left\langle V_{\omega\over 2}\left({z\over\tau},{{\bar z}\over{\bar\tau}}\right) V_{\omega\over 2}(0) \right\rangle_{{\rm Liouville}, T^2(-{1\over\tau})}=|\tau|^{4+{{\omega^2}}}\left\langle V_{\omega\over 2}(z,\bar z) V_{\omega\over 2}(0) \right\rangle_{{\rm Liouville}, T^2(\tau)},
\\
& \left\langle V_{\omega\over 2}\left({z},{{\bar z}}\right) V_{\omega\over 2}(0) \right\rangle_{{\rm Liouville}, T^2(\tau+1)}=\left\langle V_{\omega\over 2}(z,\bar z) V_{\omega\over 2}(0) \right\rangle_{{\rm Liouville}, T^2(\tau)}.
\fe
In these computations, it is important to ensure that the torus 2-point function can be expressed as a convergent expansion in either the OPE channel or the necklace channel, making use of the invariance under $z\to-z$ and $z\to z+2\pi$, as well as the $\hat q_i$ parameters introduced in section \ref{torusblock}.

There is another highly nontrivial consistency check between the OPE channel and the necklace channel, that involves analytic continuation in the energy $\omega$ of the external vertex operator. The formulae (\ref{liou_ope}) and (\ref{liou_neck}) a priori apply to real external Liouville momenta. As we analytically continue to the strip $0<{\rm Im}(\omega)<1$, no poles of the integrand cross the integration contours in $P$ and $P'$ (namely, along the real axis) in the OPE channel, or the contours in $P_1$ and $P_2$ in the necklace channel. As we continue $\omega$ such that ${\rm Im}(\omega)$ exceeds 1, in the regime $1<{\rm Im}(\omega)<2$ in particular, in the necklace channel still no poles have crossed the $P_1,P_2$ integration contour, but in the OPE channel (\ref{liou_ope}) the poles at $P=-i+\omega$ and $P=i-\omega$ have crossed the $P$ integration contour. To maintain analyticity, in the OPE channel computation we must include the integral over the original $P$-contour as in (\ref{liou_ope}), together with the residue contribution from the pole that has crossed the contour. That is to say, for $1<{\rm Im}(\omega)<2$, while (\ref{liou_neck}) remains valid, the RHS of (\ref{liou_ope}) receives an extra residue contribution
\ie{}
&-2 i\int_0^\infty { dP'\over\pi} {\rm Res}_{P\to\omega-i} {\cal C}({\omega\over 2},{\omega\over 2},P) {\cal C}(P',P',\omega-i)  {\cal F}_{\rm OPE}(1+{\omega^2\over 4}, 1+{\omega^2\over 4}, \omega^2 - 2i\omega, 1+P'^2;z,\tau) 
\\
&~~~~~~~~~~~ \times \overline{{\cal F}_{\rm OPE}}(1+{\omega^2\over 4}, 1+{\omega^2\over 4}, \omega^2 - 2i\omega, 1+P'^2;\bar z,\bar \tau) .
\label{ope_res}
\fe
Indeed we have numerically verified that the necklace channel integral (\ref{liou_neck}) agrees with the OPE channel computation taking into account the residue contribution (\ref{ope_res}).

\begin{figure}[h]
\centering
\includegraphics[width=9cm]{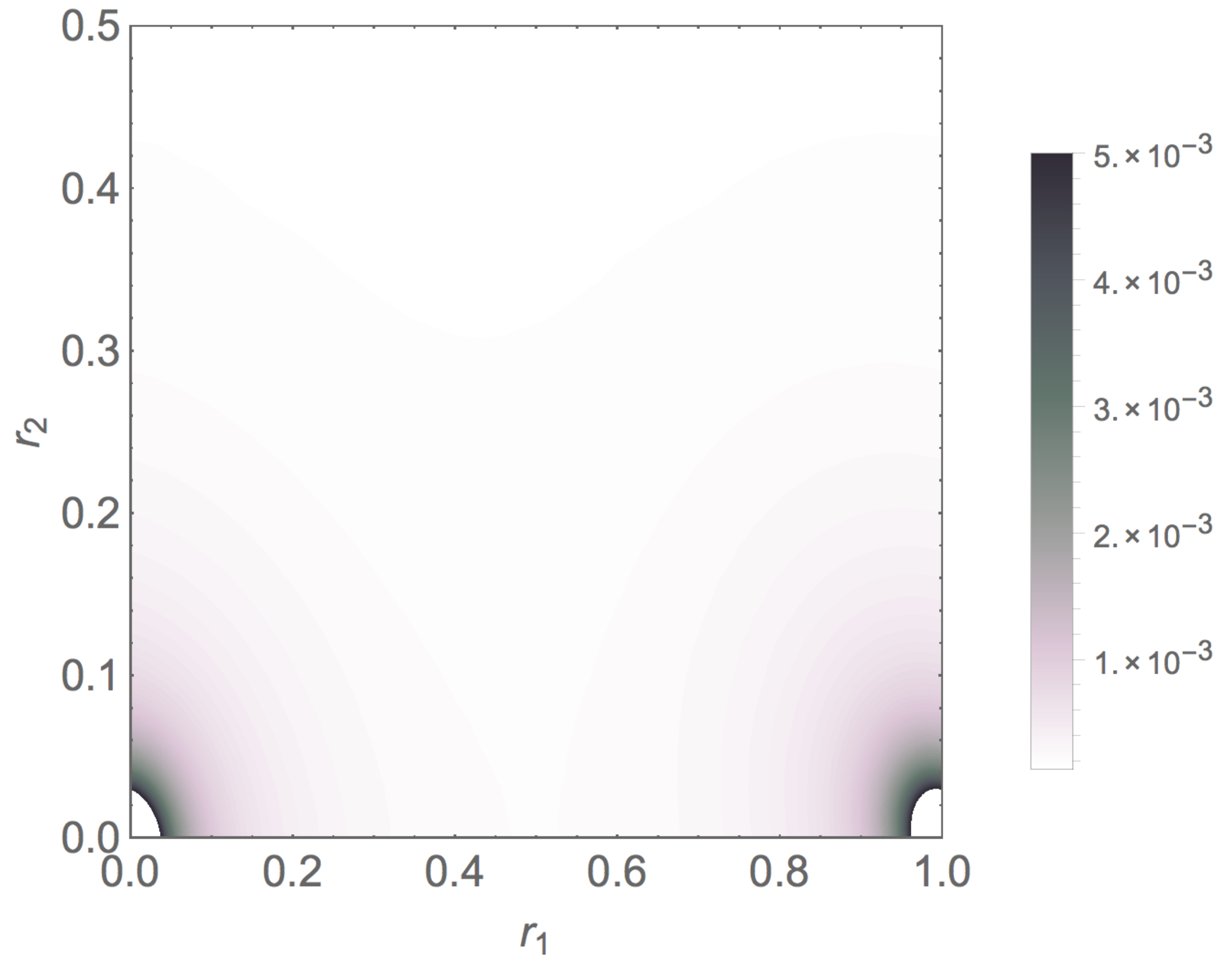}
\caption{Sample density plot of the integrand $\left|\theta_1\left({z\over 2\pi}|\tau\right) e^{-{({\rm Im} z)^2\over 4\pi \tau_2} } \right|^{\omega^2}\left\langle V_{\omega\over 2}(z,\bar z) V_{\omega\over 2}(0) \right\rangle_{{\rm Liouville}, T^2(\tau)}$ over the lower half torus, in rectangular coordinates $(r_1, r_2)$ defined by $z=2\pi(r_1+r_2\tau)$, in the case $\omega=0.8i$ and $\tau=0.25+1.25i$. This plot is produced by patching together results from the OPE channel for $|z|<2\pi \epsilon$ with $\epsilon=0.15$, and necklace channel elsewhere.}
\label{fig:toruscell}
\end{figure}

Now let us turn to the integration over the moduli space. A source of numerical error is the region of the moduli space where the integrand becomes singular, in particular near $z=0$ (or $2\pi$), where we must take care to include sufficiently many sampling points. We will also scale the sampling points accordingly in the large $\tau_2$ regime. A typical density plot of the moduli space integrand as a function of $z$ for fixed $\tau$ is shown in Figure \ref{fig:toruscell}. 
Again we cut out a small disc of radius $\delta$ around $z=0$ (or $2\pi$), and obtain the contribution from this disc by expanding the integrand in $\delta$ and performing the integral analytically. 

\begin{figure}[h]
\centering
\includegraphics[width=9cm]{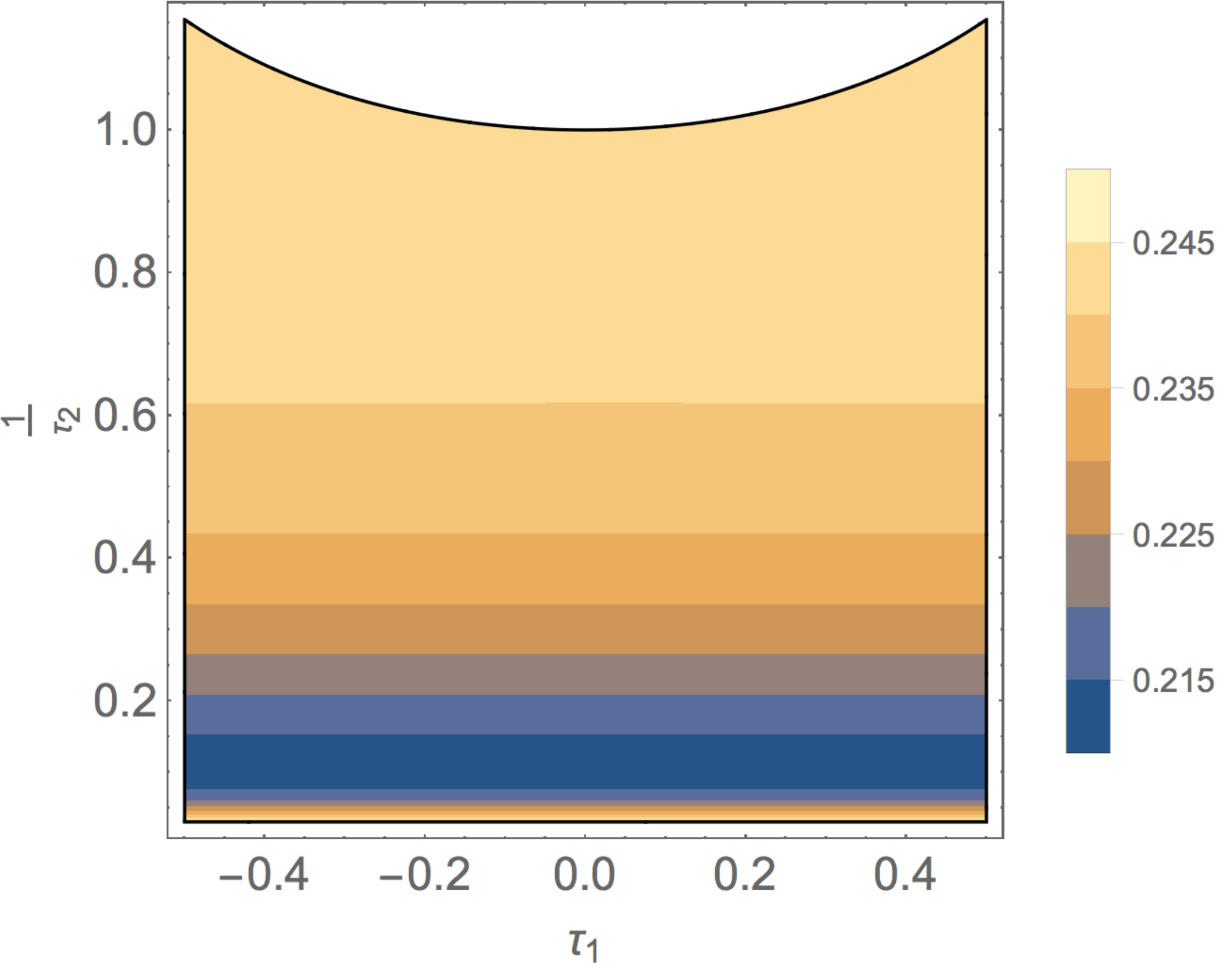}
\caption{A sample $\tau$-integrand (with integration measure $d^2\tau/\tau_2^2$) of the torus 2-point amplitude over the fundamental domain, after integration in $z$, in coordinates $(\tau_1,1/\tau_2)$, in the case $\omega=0.8i$.}
\label{fig:torusregion}
\end{figure}

After having performed the $z$-integral numerically, we have numerically verified the modular covariance in $\tau$.
Finally, to perform the $\tau$-integral over the fundamental domain, a main source of numerical error is the tail contribution near the cusp $\tau\to i\infty$. To achieve reasonable accuracy, we evaluate the $z$-integral for a set of large $\tau_2$ values (where the $\tau_1$ dependence is exponentially suppressed), and fit the result to a function in $\tau_2$ of the form $a_0\tau_2^{-2}+a_1\tau_2^{-5/3}+a_2\tau_2^{-3}$ (the leading behavior is given by (\ref{eq:largetau2})). We then integrate this fitting function over the large $\tau_2$ region of the fundamental domain. A typical plot of the $\tau$-integrand over the fundamental domain is shown in Figure \ref{fig:torusregion}.

For each fixed (imaginary) $\omega$, the integrand of (\ref{torusamp}) over the fundamental domain can be computed in parallel. The final results are shown in Figure \ref{fig:torus4imw}.

\bibliographystyle{JHEP}
\bibliography{c1}

\end{document}